\documentstyle[prb,aps,multicol,eqsecnum,epsf]{revtex}
\begin{document}
\draft
\title{Spin-dependent transport in a Luttinger liquid}
\author{L.~Balents$^1$ and R.~Egger$^{2,3}$}
\address{${}^1$~Physics Department, University of California, Santa
Barbara, CA 93106, USA\\
${}^2$~Department of Mathematics, Imperial College, 180 Queen's Gate,
London SW7 2BZ, United Kingdom\\
${}^3$~Fakult{\"a}t f{\"u}r Physik, Albert-Ludwigs-Universit{\"a}t Freiburg,
D-79104 Freiburg, Germany }
\date{Date: \today}
\maketitle
\begin{abstract}
  We develop a detailed theory for spin transport in a one-dimensional
  quantum wire described by Luttinger liquid theory.  A hydrodynamic
  description for the quantum wire is supplemented by boundary
  conditions taking into account the exchange coupling between the
  magnetization of ferromagnetic reservoirs and the boundary
  magnetization in the wire.  Spin-charge separation is shown to imply
  drastic and qualitative consequences for spin-dependent transport.
  In particular, the spin accumulation effect is quenched except for
  fine-tuned parameter regimes.  We propose several feasible setups
  involving an external magnetic field to detect this phenomenon in
  transport experiments on single-wall carbon nanotubes.  In addition,
  electron-electron backscattering processes, which do not have an
  important effect on thermodynamic properties or charge transport,
  are shown to modify spin-dependent transport through long quantum
  wires in a crucial way.
\end{abstract}

\begin{multicols}{2}

\section{Introduction}
\label{sec:intro}

Spin-polarized transport represents a new branch of mesoscopic
physics, in which both the charge and {\sl spin} of the electron are
actively manipulated.\cite{aronov,johnson,prinz,Kikkawa,GMR} In
different setups, the spin may be employed as information storage or
transport element, where the advantage over charge transport stems
from the very long spin lifetimes in many materials, and the smallness
of the dissipated power.  These advantages have already resulted in
many technological applications, and among the most popular future
perpectives of spintronics is the field of quantum
computation.\cite{qc} Spin-dependent transport also offers novel
insights into fundamental physics.  In this paper we shall address in
detail how spin transport proceeds in strongly interacting non-Fermi
liquid metals, taking the behavior of one-dimensional (1D) metals as a
paradigm in which electron-electron interactions lead to a breakdown
of Fermi liquid theory.  The 1D non-Fermi liquid behavior is often
described by {\sl Luttinger liquid}\ (LL) theory.\cite{book}

The primary motivation for this study comes from recent transport
experiments\cite{LL-tubes2} on carbon nanotubes,\cite{tubes} which
have demonstrated the breakdown of Fermi liquid theory in these nearly
ideal 1D quantum wires (QWs).  In fact, when studying charge transport
in single-wall nanotubes (SWNTs), the observed power-law behaviors in
the tunneling density of states are consistent with their theoretical
description\cite{LL-tubes1} in terms of a LL.  The LL describes metals
in the 1D limit where only one or very few bands intersect the Fermi
energy.  This non-Fermi liquid exhibits fractionalization of electrons
into novel quasiparticles, comprising a diverse set carrying spin
separately from charge, and charge in fractions of the electron charge
$e$.\cite{qp} Furthermore, the LL is the simplest model showing the
remarkable phenomenon of {\sl spin-charge separation} which has been
postulated by many to underly the cuprate
superconductors.\cite{anderson,lw,bfn,sf}\ In a LL, spin and charge
degrees of freedom are completely decoupled, and moreover
characterized by different velocities. As Landau quasiparticles are
unstable, an electron will spontaneously decay into charge and spin
density wave packets which then propagate with different velocities.
Thereby a spatial separation of spin and charge of the electron
results.  Unfortunately, this hallmark behavior of strongly correlated
1D fermions remains to be observed experimentally (at least in an
unambiguously accepted way).  Elaborating on our short
paper,\cite{PRL} we propose several feasible setups that would allow
to unambiguously detect spin-charge separation via spin transport
experiments on an individual SWNT.  For related but rather different
proposals to detect spin-charge separation in a LL via spin transport,
see also Refs.~\onlinecite{si97} and \onlinecite{si98}.

\begin{figure}
\epsfxsize=0.8\columnwidth
\epsffile{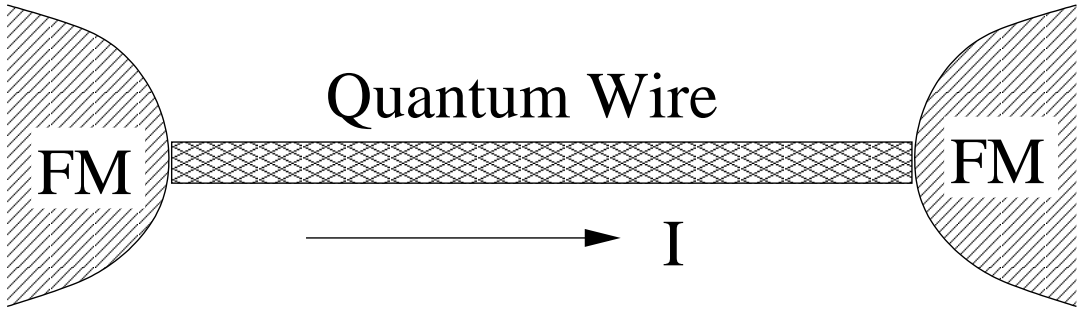}
\caption[]{\label{fig0} 
Proposed experimental setup (schematic). An individual SWNT
or quantum wire is connected via low-conductance contacts
to two ferromagnetic reservoirs, and $I-V$ curves should be measured
either in an additional magnetic field or for various angles
between the magnetization directions in the leads.
}
\end{figure}

As will be described at length below, this can be achieved by
attaching {\sl ferromagnetic}\ leads to the QW, and possibly an
additional magnetic field, see Fig.~\ref{fig0}.  For simplicity,
identical contact parameters are assumed below, with straightforward
generalization possible.  By measuring the variations of the
current-voltage $(I-V)$ 
characteristics with either the angle $\theta$ between the
ferromagnetic magnetizations in the leads, or the magnetic field
$\vec{B}$, one can indeed directly probe spin-charge separation.  A
related spin-transport experiment has been carried out recently for
multi-wall nanotubes, where the angle $\theta$ was fixed to either
zero or $\theta=\pi$.\cite{FM-tube} The experiment proposed here for
SWNTs should either allow for arbitrary $\theta$, or employ an
additional magnetic field.  We note that spin transport in such a
setup is well understood for Fermi liquids.  In particular, the $I-V$
characteristics (including a magnetic field) have been computed
recently using a semiclassical description.\cite{brataas}

In our theory, we assume that tunneling across the two
contacts proceeds incoherently, i.e., the length $L$ 
of the QW must be longer than either the thermal
length scale $\hbar v/k_B T$ or the scale $\hbar v/ eV$ set by the
applied voltage $V$.
We then consider in general systems composed of 1D
interacting quantum wires and bulk ferromagnets.  
With the exception
of Sec.~\ref{sec:extensions}, 
where the additional flavor degree of freedom present
in SWNTs will be addressed, we focus on single-mode 
QWs. 
 
Another interesting aspect of spin transport concerns the role of
electron-electron backscattering interactions.  In a spin-$\frac12$
QW, these interactions are (marginally) irrelevant under the
renormalization group (RG) flow, and therefore only cause a
renormalization of interaction parameters in the low-energy LL
theory.\cite{book} However, this essentially thermodynamic
(equilibrium) argument must be re-examined when dealing
with spin transport.  In fact, such interactions, despite
being irrelevant, can result in nonlinear and sometimes dramatic
effects, e.g., a non-sinusoidal oscillatory $I-V$ characteristics.

The structure of this paper is as follows.
In Sec.~\ref{sec:formulation}, we introduce the
basic model and outline the computation of the
non-equilibrium spin current.
In Sec.~\ref{sec:contacts}, the physics arising
at a contact between a ferromagnetic reservoir 
and a Luttinger liquid is addressed at length.
Two processes are shown to be of importance,
namely electron tunneling and boundary exchange.
Exchange leads to novel conformally-invariant
boundary conditions which are derived here.
In Sec.~\ref{sec:bulk}, a hydrodynamic description
of spin transport in the 1D QW is developed.
In Sec.~\ref{sec:app1}, we derive the 
$I-V$ characteristics in a magnetic
field for the simplest spin transport device, 
see Fig.~\ref{fig0}, first for a short-to-intermediate 
length of the QW.  Under the latter condition,
backscattering can be neglected.  The effects
of backscattering are then addressed in detail
in Secs.~\ref{sec:cc} and \ref{sec:dd}, where we focus on zero
magnetic field for clarity.  Finally, several
extensions and possible concerns are addressed
in Sec.~\ref{sec:extensions}. We conclude
in Sec.~\ref{sec:outlook} by discussing an
analogy to ballistic superconductor-normal-superconductor
(SNS) junctions, summarizing some
open questions and providing an outlook.
Details of our calculations in Sec.~\ref{sec:app1}
can be found in three appendices.
In intermediate steps of the calculations,
we put $e=\hbar=1$, but restore units in
experimentally relevant results.

\section{Model and Formulation}
\label{sec:formulation} 

The low-energy 
description of a single-mode QW is remarkably universal,\cite{book}
and a sufficient Hamiltonian for our purposes is 
\begin{equation} \label{Lwire}
  H_{\rm QW} = \int_{0}^L \!\!\!\!dx\, \left\{
    -i \psi^\dagger
      v \tau^z \partial_x\psi^{\vphantom\dagger} + u
    \left(\psi^\dagger\psi^{\vphantom\dagger}\right)^2 \right\} \;,
\end{equation}
where $\psi = \psi_{a\alpha}$ is a four-component spinor.  Here
$a=R/L$ indexes the chirality that differentiates right- and
left-moving modes, and $\alpha=\uparrow/\downarrow$ indexes the spin.
We suppress the indices whenever possible, employing Pauli matrices
$\vec\tau$ and $\vec\sigma$ acting in the chirality and spin spaces,
respectively.  Furthermore, $v$ denotes the Fermi velocity, or, more
generally, the spin velocity of the interacting theory.  If
the QW is isolated, the zero-current condition at the endpoints
requires to impose the boundary conditions $\psi_R(0) = \psi_L(0)$ and
$\psi_R(L)=\psi_L(L)$.  Only the forward-scattering interaction $u$ is
kept in Eq.~(\ref{Lwire}).  Alternatively, we will use the exponent
$\alpha>0$ for tunneling into the end of the LL, e.g., at $x=0$, as a
measure of the interaction strength.  Eq.~(\ref{Lwire}) is not
completely general.  It contains two parameters, $v$ and $u$, while a
general single-mode LL has three parameters: a charge velocity $v_c$,
a spin velocity $v$, and a dimensionless ``Luttinger parameter'',
often denoted $K_\rho$ or $g$, where $\alpha = (K_\rho^{-1}-1)/2$.
In Eq.~(\ref{Lwire}), we have assumed  full (Galilean) translational
invariance, leading to $K_\rho = v/v_c$.  This relation is 
expected to be well-satisfied in many experimentally relevant QWs, and
moreover the manipulations to follow relax this condition and thus can
be applied even when $K_\rho \neq v/v_c$.

In some circumstances,
Eq.~(\ref{Lwire}) should be supplemented by the electron-electron
{\sl backscattering} interaction, 
\begin{equation} \label{pert}
H_{\rm bs} = - bv \int dx \vec{J}_L \cdot \vec{J}_R \;,
\end{equation}
with the chiral spin currents 
\begin{equation}
\vec{J}_{R/L} (x) =  \frac12 :\psi^\dagger_{R/L} (x)\vec{\sigma} 
\psi^{\vphantom\dagger}_{R/L} (x): \;,
\end{equation}
where the colons denote normal ordering.  These chiral spin currents
obey Kac-Moody commutation relations ($\mu,\nu=x,y,z$),
\begin{eqnarray} \label{km}
  [J_{L/R}^\mu(x),J_{L/R}^\nu(x')]  &=&
 \pm i\delta'(x-x')\delta^{\mu\nu} \\ \nonumber &+& i 
  \epsilon^{\mu\nu\lambda} J_{L/R}^\lambda(x) \delta(x-x') \;,
\end{eqnarray}
where the $+$ ($-$) sign is associated with the $L$ ($R$)
current.   We note that in a spin-$1/2$ QW, the
backscattering interaction (\ref{pert}) is  marginally 
irrelevant in the RG sense, and hence can be
neglected at low energies in many equilibrium properties.
In a SWNT, the generalization of $H_{\rm bs}$ causes exponentially small gaps 
that can be neglected at not too low energies.
Furthermore, the dimensionless
backscattering coupling constant $b$ is generally small
and scales as $1/R$ with the tube radius $R$. 
Importantly, as will be discussed below in detail,
a precession effect encoded in Eq.~(\ref{pert}) 
is crucial for understanding spin transport in long QWs.

For energies well below the electronic bandwidth $D$, a ferromagnetic
(FM) lead can be described using an effectively non-interacting Stoner-like
picture\cite{fazekas} with a constant density of states.
It is then sufficient to employ a non-interacting 1D model,
e.g., for the left lead ($x<0$),
\begin{equation} \label{LFM}
  H_{\rm FM} = \sum_{s=\pm 1} \int_{-\infty}^0 \!\!\!\!dx\, 
    f^\dagger\left(-iv_s \hat{u}_s \tau^z
      \partial_x\right)f^{\vphantom\dagger}\;,  
\end{equation}
where $f$ is again a four-component spinor.
Comparing to  Eq.~(\ref{Lwire}), different
spin quantization axes have been used, and therefore
the projection operator 
\begin{equation} \label{pop}
\hat{u}_s = (1\pm \hat{m}\cdot\vec\sigma)/2 
\end{equation}
 projecting the spin quantization
axis of the QW onto the magnetization $\hat{m}$ is needed.
This description of the semi-infinite lead must be supplemented by an
appropriate boundary condition,  $f_R(0) = f_L(0)$.
In Eq.~(\ref{LFM}), the two Fermi velocities $v_\pm$ parameterize the
different densities of states, 
$\rho_s=1/(2\pi v_s)$, for the majority and
minority carriers.  Following Ref.~\onlinecite{brataas}, we choose a suitable
rescaling of the $f$ operators 
to set $v_+ = v_- = 1$, thereby incorporating
the difference in the density of states into a redefinition of the
hopping matrix elements, $t_s \rightarrow t_s/\rho_s$, employed in 
the tunneling Hamiltonian, see Eq.~(\ref{tun}) below.
Formally this is done by choosing eigenstates $f_s$ of
$\hat{m}\cdot\vec{\sigma}$ with eigenvalue $s=\pm$, and then rescaling
$f_s(x) \rightarrow v_s^{-1} f_s(x/v_s)$, the spatial rescaling being
allowed because the different spin polarizations are non-interacting and
tunneling acts only at $x=0$.  

The LLs and FMs in question will be considered coupled by
low-conductance contacts with identical properties. 
Processes in which electrons are transferred across such a 
contact can be described by the {\sl tunneling
Hamiltonian} $H_{\rm tun}$.  Provided this contact occurs at one of the
ends  of the LL, say, at $x=0$, this has the form
\begin{equation} \label{tun}
  H_{\rm tun} = F^\dagger W \Psi^{\vphantom\dagger} + \Psi^\dagger
  W^\dagger F^{\vphantom\dagger}\;, 
\end{equation}
where $F=f(0^-)$ and $\Psi=\psi(0^+)$ 
are Fermion annihilation operators at the ends of the
FM and QW, respectively.  The 2$\times$2 tunneling
matrix $W$ reads with Eq.~(\ref{pop}),
\begin{equation}
W = \sum_{s=\pm 1} t_s \hat{u}_s \;.
\end{equation}
Using the spin-dependent hopping matrix elements $t_s$, 
we may define spin-dependent conductances $G_s
= (e^2/h) |t_s|^2$, or, alternatively, the contact
parameters
\begin{equation}  \label{junctpar}
G=G_\uparrow+G_\downarrow \;, \quad P = (G_\uparrow-G_\downarrow)/G \;.
\end{equation}
Here $G$ is the total conductance associated with the contact.  In a
slight abuse of terminology, we call $P$ the {\sl polarization}.  The
polarization satisfies $0\leq P \leq 1$, and in fact represents the
asymmetry between the {\sl local} tunneling density of states of the
majority and minority carriers of the ferromagnet.  Hence $P$ is not a
bulk property, and depends upon the detailed nature of the FM-QW
contact.  In the experiment of
Ref.~\onlinecite{FM-tube}, application of our theoretical results, in
particular Eq.~(\ref{res1}) -- which differs slightly from the theory
used in Ref.~\onlinecite{FM-tube} -- gives a value of $P=0.3$ for a
multi-wall NT to FM (cobalt) contact at $T=4.2K$.

$ $From Eq.~(\ref{tun}), one may deduce the
spin current, defined by 
\begin{equation}
  \vec{J}^{\rm tun} = {\partial \over {\partial t}} \left.\left( \int\! dx\, 
    \vec{M} \right)\right|_{\rm tun} \;,
\end{equation}
where  the magnetization density in the QW is
\begin{equation}
  \vec{M} = \vec{J}_R + \vec{J}_L \;.
\end{equation}
Using the continuity equation
in the absence of backscattering interactions, $b=0$, the steady-state
current is thus
\begin{equation}\label{defsc}
 \vec{J} = v (\vec{J}_R - \vec{J}_L) \;.
\end{equation}
Then the tunneling current is
\begin{equation} \label{spincurrent}
  \vec{J}_{\rm tun} = {i \over 2} \left(F^\dagger W \vec\sigma
    \Psi^{\vphantom\dagger} - \Psi^\dagger \vec\sigma \, W^\dagger
    F^{\vphantom\dagger} \right)\;. 
\end{equation}
Of course, a formula similar to Eq.~(\ref{spincurrent}) obtains for the
charge current across the contact,
\begin{equation} \label{chargecurrent}
  I = i\left(F^\dagger W \Psi^{\vphantom\dagger} - \Psi^\dagger W^\dagger
    F^{\vphantom\dagger}\right)\;. 
\end{equation}
In addition to the tunnel coupling in Eq.~(\ref{tun}), the FM 
magnetization and the LL boundary magnetization
can be coupled by a pure {\sl exchange} term,
\begin{equation} \label{K1}
  H_{\rm ex} = - K \hat{m}\cdot \Psi^\dagger {\vec\sigma \over 2}
  \Psi^{\vphantom\dagger}\;. 
\end{equation}
Even if in a microscopic formulation no bare exchange coupling $K$ is
present, it will be generated in the low-energy
effective Hamiltonian since tunneling causes
virtual processes corresponding to exchange, 
see below and Ref.~\onlinecite{slonc}.

To study transport, we must formulate the non-equilibrium dynamics of
the system.  This formulation is more subtle than in conventional
charge transport, due to the complications arising from the
non-commuting nature of spin.  We therefore proceed carefully along
the lines of a Keldysh approach.  For concreteness, we specialize for
the moment to a semi-infinite  LL contacted at $x=0$.
Consider an initial system composed
of two decoupled pieces, described by the
Hamiltonian $H_0 = H_{\rm FM} + H_{\rm QW}$.  The ferromagnetic part,
governed by $H_{\rm FM}$, is polarized along direction $\hat{m}$, and
located at $x<0$.  The wire, located at $x>0$, is governed by
$H_{\rm QW}$, which is assumed $SU(2)$ invariant.  The latter condition
guarantees the existence of a continuity equation for the spin density
and current.  Similarly, charge conservation implies a continuity
equation for charge density and current.  At $t=-\infty$, we assume
that each half is at quasi-equilibrium at its own chemical potential,
$\mu_{\rm FM}$ and $\mu_{\rm QW}$.  
Similarly, we assume that the LL supports a 
quasi-equilibrium magnetization, which can be described by a grand canonical 
distribution with a ``spin chemical potential'' $\vec{h}$.  
We stress that neither $\mu_{\rm QW}$ nor $\vec{h}$ are physical potentials,
such as electrostatic or Zeeman
fields, but rather characterize the initial non-equilibrium
distribution.  Then we adiabatically turn 
on the contact perturbation $H' = H_{\rm 
  tun} + H_{\rm ex}$,
\begin{equation}
  H(t) = H_0 + e^{\delta t} H^\prime\;,
\end{equation}
where $\delta \rightarrow 0^+$ is an infinitesimal inverse time-scale
controlling the slow turning on of the contact interaction. 

The above formulation is rather rigorous, but has the limitation of
being formulated to treat {\sl both}\ the tunneling and boundary
exchange terms as perturbations.  In practice, this is appropriate in
the low-conductance limit, since the boundary exchange is generally
determined by virtual tunneling processes, and is hence small.
However, it is useful theoretically to contemplate a situation in
which the tunneling is small but the boundary exchange is not.  In
such a case, it appears natural to include $H_{\rm ex}$ into $H_0$
rather than in $H'$.  This raises difficult conceptual issues, since
$H_{\rm ex}$ does not commute with the total spin of the LL, and hence
renders its magnetization uncertain in directions perpendicular to
$\hat{m}$.  On physical grounds, however, we expect that, because
$H_{\rm ex}$ transfers spin into/out of the LL only at the boundary,
sense can be made of the {\sl bulk} spin chemical potential in the
limit of a semi-infinite LL.  This is fairly clear from the following
thought experiment: imagine preparing the LL with zero tunneling and
zero boundary exchange in a state with a non-zero
magnetization density not parallel to $\hat{m}$.  Then, if $H_{\rm ex}$ 
is turned on at some time, its effect will be to scatter
left-moving electrons into right-moving ones upon their reaching
$x=0$, changing their spin orientation in the process.  If the LL is
semi-infinite, however, it would take an infinite amount of time to
modify the mean magnetization of the LL in this way.  Instead, one
expects a steady-state to be established, generally with a
time-independent spin current.  Moreover, for a finite but long LL, so
long as there are some inelastic processes deep in the LL that can
equilibrate the returning electrons, one expects that an equilibrium
state will be established which has a mean magnetization very close to
that before turning on the boundary exchange.  In fact, a scattering
approach of this type can be directly implemented using bosonization
methods, and will be discussed analytically in Sec.~\ref{sec:bos}.
Alternatively, $H_{\rm ex}$ can be incorporated directly into $H_0$,
but in this case, care must be taken to ensure that $\vec{h}$ is
coupled only to the magnetization outside a neighborhood of the
boundary.  For the moment, we simply use the above discussion as
motivation to incorporate $H_{\rm ex}$ into $H_0$.

We are interested in the properties of the system at time $t=0$, when
a steady-state transport of charge and spin between the two systems
has been achieved.  Then we can formally calculate the expectation
value of any operator at this time:
\begin{eqnarray}
  \langle {\cal O}(0) \rangle & = & {1 \over Z} {\rm Tr}\, \bigg[
    e^{-\beta (H_0 - \Lambda)} {\cal T} \exp \left( i \int_{-\infty}^0 dt
      H(t)\right) {\cal O} \nonumber \\
    & & \times {\cal T} \exp \left( -i \int_{-\infty}^0 dt
      H(t)\right) \bigg] \;,
\end{eqnarray}
where ${\cal T}$ denotes time ordering and
\begin{equation}
  \Lambda = \mu_{\rm FM} N_{\rm FM} + \mu_{\rm QW} N_{\rm QW} +
\vec{h}\cdot \vec{S}_{\rm QW}^{\rm tot} \;.
\end{equation}
It will be most convenient to choose the zero of energy in $H$ such
that $\mu_{\rm FM} = \mu_{\rm QW} = h = 0$ in equilibrium, i.e.,
for zero applied voltage.

Expanding out the time-ordered exponential, one obtains to lowest order
\begin{equation} \label{expect}
  \langle {\cal O}\rangle = -i \int_{-\infty}^0 \! dt\, {1 \over Z} 
  {\rm Tr}\, \left( e^{-\beta \tilde{H}_0} \left[ {\cal
        O},H'(t)\right] \right)e^{\delta t}\;, 
\end{equation}
where $\tilde{H}_0 = H_0 - \Lambda$  and $H'(t) = e^{iH_0 t} H'
e^{-iH_0 t}$.  Owing to the factor of $\Lambda$ in the Boltzmann
weight, not present in the time evolution of $H'(t)$, this is a
non-equilibrium expectation value.  We can cast it into a more equilibrium 
form by writing $H_0 = \tilde{H}_0 + \Lambda$.  Then
\begin{equation} \label{rewrite}
  H'(t) = e^{i\tilde{H}_0 t} \left( e^{i\Lambda t} H' e^{-i\Lambda
      t}\right) e^{-i\tilde{H}_0 t} \;. 
\end{equation}
Equation (\ref{rewrite}) has the form of the time-evolution of the operator 
in the brackets evolved by the ``non-equilibrium'' Hamiltonian
$\tilde{H}_0$, which is the same as is used in the Boltzmann weight in 
Eq.~(\ref{expect}).
Equation (\ref{expect}) can be rewritten in the slightly more suggestive
form,
\begin{equation} \label{expect2} 
  \langle {\cal O}\rangle = -i \int_{-\infty}^0 \! dt\, 
e^{\delta t} \left\langle
    \left[ {\cal O},H'_\Lambda(t)\right] 
\right\rangle_{\tilde{H}_0}\;,
\end{equation}
where $H'_\Lambda$ indicates the modified operator
\begin{equation} \label{hlambda}
  H'_\Lambda = e^{i\Lambda t} H' e^{-i\Lambda t}\;,
\end{equation}
which evolves according to the fictitious equilibrium time evolution
dictated by $\tilde{H}_0$. The subscript $\tilde{H}_0$ on the
expectation value indicates that it is a standard equilibrium average
with respect to the Hamiltonian $\tilde{H}_0$, in which the argument
of the Fermion fields indicates standard Heisenberg picture time
dependence using $\tilde{H}_0$, 
\begin{eqnarray*}
  \langle {\cal O}_1\cdots{\cal O}_n \rangle_{\tilde{H}_0} & = & {1 \over
    Z} {\rm Tr}\, \left(e^{-\beta\tilde{H}_0} {\cal O}_1\cdots{\cal
      O}_n \right) \;, \\
  {\cal O}(t) & = & e^{i\tilde{H}_0 t} {\cal O} e^{-i\tilde{H}_0 t}\;.
\end{eqnarray*}
Thereby we can express an intrinsically
non-equilibrium property of the system with Hamiltonian $H_0$ in terms 
of a fictitious equilibrium average with respect to the shifted
Hamiltonian $\tilde{H}_0$.  

\section{Contacts}
\label{sec:contacts}

In this section, we analyze in detail
the physics of a single contact between a
FM lead and a semi-infinite LL (taken at $x=0$).  
We expect that the results
apply to finite-length LLs longer than the thermal length 
$v/k_B T$ beyond which transport is incoherent.

\subsection{End contacts: Boundary operators}

We first study the properties of the contact in equilibrium from the
RG point of view. It is helpful to view
both $H_{\rm tun}$ and $H_{\rm ex}$ as perturbations to a decoupled
fixed point described by $H_0$.  Standard arguments 
give the scaling dimension of both $t_s$ and $K$, 
\begin{equation}
\Delta_{t_s} = 1+ \alpha/2 \;, \quad 
\Delta_K = 1  \;.
\end{equation} 
The scaling dimension $\Delta_K$ is not renormalized
due to spin-charge separation in the QW.   A simple
calculation then gives the RG scaling equations
\begin{equation} \label{RGeqns}
  \partial_\ell|t_s|^2(\ell)  = -\alpha |t_s|^2 \;, 
\qquad \partial_\ell K(\ell)= c\left( |t_\uparrow|^2 -
    |t_\downarrow|^2\right) \;, 
\end{equation}
where $\ell = \ln (D/E)$ is the standard RG flow parameter, 
and $c$ denotes a non-universal constant. 
Note that $K$ is renormalized by the hopping matrix elements,
but the converse does not occur.  
The most important property of Eq.~(\ref{RGeqns}) is that
while the tunneling is {\sl irrelevant}\ for $\alpha>0$, 
the exchange coupling $K$
between the LL and FM is exactly {\sl marginal}.  

Following the RG flow from the ultraviolet cutoff $D$ down to 
energy $E\approx {\rm max}(k_B T, eV) \ll D$, we find $|t_s|^2(E) = 
|t_s|^2 (E/D)^\alpha$, and
\begin{equation}\label{RGsol}
K(E)  =  K + \alpha^{-1} c GP [1-(E/D)^\alpha] \gg |t_s|^2(E)\;.
\end{equation}
Therefore the effective exchange coupling $K(E)$ does not pick up the 
$(E/D)^\alpha$ suppression factor, and generally is much larger
than the effective hopping $t_s(E)$, regardless of the microscopic
``bare'' values of these couplings.
Parenthetically, we note that the non-interacting limit, $\alpha\to 0$,
is not correctly handled by the simple RG equations (\ref{RGeqns}),
as this limit involves additional marginal operators.
In fact, for $\alpha\to 0$, not logarithmic dependencies
[as predicted by Eq.~(\ref{RGsol})] but instead
a principal parts prescription emerges.  

Since the exchange coupling is exactly marginal and the
tunneling irrelevant, solving the problem for {\sl zero tunneling} but 
non-zero exchange gives the ``boundary fixed point'' solution.  From
the viewpoint of low-energy physics, the only effect of the boundary exchange
coupling is then to induce a modified boundary condition at the contact.
This modified conformally invariant boundary condition 
comprises a novel boundary fixed
point\cite{cardy} describing the semi-infinite LL 
close to a ferromagnet.
 
\subsection{Zero Tunneling Boundary Fixed Point}
\label{sec:bos}

To gain maximum insight into the physics, we solve the equilibrium
problem with zero tunneling exactly in a number of equivalent ways.
The most familiar method is Abelian bosonization.\cite{book}
Choosing the quantization axis for the spinor basis along the
$\hat{m}$ axis, the electron field can be written in terms of boson
fields,
\begin{equation}
 \sum_{s'}  \hat{m}\cdot\vec\sigma_{ss'}\psi_{R/L,s'}
   =  \frac{1}{\sqrt{a_0} }
\, e^{{i \over \sqrt{2}}[\varphi_\rho \pm \theta_\rho +
    s(\varphi_\sigma \pm \theta_\sigma)]}  \;,
\end{equation}
where $(\varphi_\rho,\theta_\rho)$ and $(\varphi_\sigma,\theta_\sigma)$ are
charge and spin bosons, respectively, that satisfy
the algebra
\[
   [\theta_{\rho/\sigma}(x), \varphi_{\rho/\sigma}(x')] = i\pi \Theta(x-x')
\]
with the Heaviside step function $\Theta(x)$.  The short-distance cutoff
$a_0$ describes a non-universal scale factor
relating the microscopic Fermion field to the continuum bosonized
vertex operators and is related to the bandwidth, $a_0\simeq v/D$.

Neglecting the bulk backscattering interaction (\ref{pert}) 
for the moment, the LL Hamiltonian splits into spin and charge 
components, $H=H_\rho +
H_\sigma$.  We require only the spin component,
\begin{equation} \label{spinboson}
  H_\sigma = {v \over {2\pi}} \int_0^\infty \!dx\,  \left[(\partial_x
      \theta_\sigma)^2 + (\partial_x \varphi_\sigma)^2 \right]  -
    {K \over  {\pi\sqrt{2}}} \partial_x\theta_\sigma (0) \;.
\end{equation}
The last term is the boundary exchange term.  It can be
transformed away by the canonical transformation
\begin{equation} \label{shift}
  \theta_\sigma(x) \rightarrow \theta_\sigma(x) + {K \over
    {v \sqrt{2}}} \Theta(x-0^+)\;,
\end{equation}
which simultaneously encapsulates several physical effects.  First, since
\[
\vec{S}\cdot\hat{m} \simeq \partial_x\theta_\sigma/(\pi\sqrt{2}) \;,
\]
a local ``proximity effect'' magnetization is induced 
in the neighborhood of the contact.  As this 
induced magnetization decays on the microscopic scale of the 
Fermi wavelength, the magnetization appears within bosonization as a
delta function.  Second, the {\sl transverse} left- and right-moving spin
currents also depend on $\theta_\sigma$,
\[
  J_R^\pm \sim  e^{\pm i\sqrt{2}(\varphi_\sigma+\theta_\sigma)} \;, \quad
  J_L^\pm \sim  e^{\pm i\sqrt{2}(\varphi_\sigma-\theta_\sigma)} \;.
\]
Thus the shift in Eq.~(\ref{shift}) leads to modifications of the 
transverse spin current,
\begin{eqnarray}
  J_R^\pm(x>0^+) & \rightarrow & e^{\pm i \vartheta/2} J_R^\pm(x>0^+)\;, \\
  J_L^\pm(x>0^+) & \rightarrow & e^{\mp i \vartheta/2} J_L^\pm(x>0^+)\;,
\end{eqnarray}
where a dimensionless measure of the boundary exchange coupling
is provided by the {\sl exchange angle},
\begin{equation}
  \vartheta =2K/ v \;.
\end{equation}
This transformation can thus be
interpreted physically as a {\sl phase shift}.  Where before the
transformation spin conservation required $\vec{J}_R(0^+) =
\vec{J}_L(0^+)$, after the change of variables, we have
\begin{equation}
  J_R^\pm(0^+) = e^{\pm i \vartheta} J_L^\pm (0^+) \;.
\end{equation}
Unlike the purely local magnetization parallel to $\hat{m}$, this
phase shift can have measurable consequences far from the contact.  

The phase shift can also be understood directly in terms of electrons, 
which is useful for making contact with earlier non-interacting
theories.\cite{brataas} It is simplest to combine the right- 
and left-moving Fermions of the semi-infinite LL into a single
  chiral right-moving Fermion for each spin species on the full
infinite line.  In particular, we let 
\begin{equation}
  \psi'_\alpha(x) = \cases{\psi_{L\alpha}(-x) & $(x<0)$ \cr
    \psi_{R\alpha}(x) & $(x>0)$}\;,
\end{equation}
which ensures continuity of $\psi'$ at the origin due to the boundary
condition $\psi_R(0)=\psi_L(0)$.  At the boundary where the original
right- and left-movers have been ``merged'' together, we have
$\vec{J}_R(0)=\vec{J}_L(0)=\vec{J}'(0)$, so that the exchange
Hamiltonian becomes
\begin{equation} \label{chiralexchange}
  H_{\rm ex} = - 2 K \hat{m} \cdot \vec{J}'(x=0)\;, 
\end{equation}
with $\vec{J}' = (\psi')^\dagger \vec\sigma \psi'/2$.
We may then write down the Dirac equation for $\psi'$ as
\begin{equation} \label{chiral_schroedinger}
  (\partial_t + v\partial_x)\psi'(x) = i K \delta(x)
  \hat{m}\cdot\vec\sigma \psi'(x) \;.   
\end{equation}
For low-energy stationary states, the time derivative may be
neglected, and Eq.~(\ref{chiral_schroedinger}) then yields
\begin{equation}
  \psi'(x=0^+) = \exp \left(i {\vartheta \over 2} \hat{m}\cdot\vec\sigma
\right) \psi'(x=0^-)\;. 
\end{equation}
Thus the boundary exchange simply induces different phase shifts for
(left-moving) electrons incident upon the contact from the LL and
reflected back into the LL (as right-movers), dependent upon their
polarization relative to $\hat{m}$.  

It is technically most convenient to work directly with the spin
currents.  This has the advantage of keeping the spin quantization
axis arbitrary at all stages.  Using the same ``merged'' operators
defined above, the Kac-Moody commutation
relations (\ref{km}) and Eq.~(\ref{chiralexchange}) result in the equation
of motion for the merged chiral spin current, 
\begin{equation} \label{edgeeom}
  (\partial_t+v\partial_x)\vec{J}' = 2K \delta(x) \hat{m} \times
  \vec{J}' \;,
\end{equation}
where bulk backscattering is  again neglected.
In a steady state, $\partial_t \vec{J}' =0$, and Eq.~(\ref{edgeeom})
can be formally solved to obtain
\begin{equation}\label{bcc}
\vec{J}_R(0^+) = {\cal R}(\vartheta)\vec{J}_L(0^+) \;.
\end{equation}
Here the phase shift is encoded in the
one-parameter $SO(3)$ matrix, 
\begin{equation} \label{so3}
{\cal R}(\vartheta) = \exp(\vartheta \Gamma) \;,\quad
\Gamma_{\mu\nu}=\sum_\lambda \hat{m}_\lambda \epsilon_{\lambda\mu\nu} \;,
\end{equation}
which describes rotation by an angle $\vartheta$ around the rotation
axis $\hat{m}$.  Equations (\ref{bcc}) and (\ref{so3}) 
provide the most general
formulation of the effects of boundary exchange.  Using them, we take
$\vartheta$ to {\sl define} the dimensionless ``exchange coupling
constant'' of the low-energy theory.  It describes the angle that an
incident spin in the LL precesses around the FM magnetization
direction $\hat{m}$ due to the exchange interaction.  In principle,
since the boundary exchange operator is exactly marginal, $\vartheta$
need not be small, but for the case of low-conductance contacts, one
has $\vartheta\ll 1$, see Ref.~\onlinecite{slonc}.  The boundary
condition (\ref{bcc}) describes the zero tunneling boundary fixed
point in the presence of exchange and is crucial to the subsequent
development of our theory. 

\subsection{General formulation}

We now include the effect of tunneling on top of the boundary exchange 
above.  To do this is slightly subtle, owing to an (unphysical)
short-distance singularity inherent to the linearized spectrum of the
Luttinger model.  To resolve this difficulty, we are required to
choose some short-distance regularization for the microscopic physics
of the contact.  The form of the resulting macroscopic equations 
is independent of this choice, although the quantitative values
of certain $O(1)$ coefficients can be cut-off dependent.  A convenient 
method is to employ the combined infinite chiral Fermion description
introduced above, and then assume that the tunneling occurs on the
right-moving branch slightly after 
the exchange coupling acts, i.e., within some distance
of the order of $a_0$.  When an electron tunnels
into the LL from the FM, its spin and charge are propagated to the
right and do not themselves interact with the exchange torque, so that
\begin{equation}
\vec{J}_L  = \vec{J}_L(0^+)\;, \quad
\vec{J}_R  = \vec{J}_R(0^+) +  \frac{1}{v} \vec{J}_{\rm tun} \;.
\end{equation}
The additional tunneling spin current, $\vec{J}_{\rm tun}$, can 
now be calculated using the time-dependent perturbation theory treatment
described in Sec.~\ref{sec:formulation}.

In particular, we consider ${\cal O} = \vec{J}_{\rm tun}$
with Eq.~(\ref{spincurrent}), and $H' = H_{\rm 
  tun}$.  For this case, Eq.~(\ref{hlambda}) yields
$H'_\Lambda =  A^{\vphantom\dagger}(t) + A^\dagger(t)$,
where
\[
  A(t) = F^\dagger   W U(t) \Psi^{\vphantom\dagger} \;,
\]
with the unitary matrix 
\begin{equation}
  U(t) = \exp\left[ i\left(V+ 
    {{\vec{h}\cdot\vec\sigma} \over 2}\right)t\right]\;,
\end{equation}
where $V=\mu_{\rm QW}-\mu_{\rm FM}$.
Note that $U(t)$ is simply a matrix, and hence does not represent
an operator in the Hilbert space.  It comprises the only explicit
time-dependence in the integrand in Eq.~(\ref{expect2}), and can be
removed outside the trace.  Applying the above results to
Eq.~(\ref{expect2}), we find (repeated indices are summed)
\begin{eqnarray} 
 \langle \vec{J}_{\rm tun}\rangle & = & {\rm Re}\, \int_{-\infty}^0 \! dt\,
 e^{\delta t} \left(W\vec\sigma\right)_{\alpha\beta}
 \left(U^\dagger(t)W^\dagger\right)_{\gamma\lambda}  \nonumber \\
\label{Jformula}
 &\times& \left\langle
    \left[F_\alpha^\dagger(0)\Psi_\beta^{\vphantom\dagger}(0)  ,
      \Psi_\gamma^\dagger(t),F_\lambda^{\vphantom\dagger}(t) 
      \right]\right\rangle_{\tilde{H}_0} \;. 
\end{eqnarray}
A formula similar to Eq.~(\ref{Jformula}) obtains for the
charge current (\ref{chargecurrent}),
\begin{eqnarray}
 \langle I \rangle & = & 2 \,{\rm Re}\,\int_{-\infty}^0 \! dt\,
 e^{\delta t} W_{\alpha\beta}
 \left(U^\dagger(t)W^\dagger\right)_{\gamma\lambda}  \nonumber \\
 &\times& \left\langle
    \left[ F_\alpha^\dagger(0)\Psi_\beta^{\vphantom\dagger}(0),
      \Psi_\gamma^\dagger(t)F_\lambda^{\vphantom\dagger}(t) 
      \right] \right\rangle_{\tilde{H}_0} \;. \label{Iformula}
\end{eqnarray}
Thereby both the charge and spin current across the contact can be
calculated in terms of equilibrium correlation functions.

To calculate these correlation functions, it is more convenient to
switch to a Euclidean Lagrangian approach.  Note that we require
correlators calculated not with 
respect to $H_0$, but to $\tilde{H}_0$.  Thus we must consider
\begin{eqnarray*}
  \tilde{L}_{\rm FM} & = & L_{\rm FM} - \int\! 
d\tau\int_{-\infty}^0 \!\!\!\! dx\,
  \mu_{\rm FM} f^\dagger f^{\vphantom\dagger}\;,  \\
  \tilde{L}_{\rm QW} & = & L_{\rm QW} -
  \int\! d\tau\int_0^\infty\!\!\!\! dx\, \left( \mu_{\rm QW} \psi^\dagger
    \psi^{\vphantom\dagger} + 
  \vec{h}\cdot \psi^\dagger {{\vec\sigma} \over 2}
  \psi^{\vphantom\dagger} \right) \;. 
\end{eqnarray*}
Fortunately, this modification of the Lagrangians has  no effect
on the $F$ and $\Psi$ correlators.  Physically, this is because the
lead and wire correlators are calculated in equilibrium, so that 
the added terms act as potentials.
 They thus simply rigidly shift the spectrum of states
on both sides of the contact, and those states raised or lowered below the
(equilibrium) chemical potential are filled or emptied, respectively.
In general, this would induce some weak change in the correlators due
to energy dependence of the density of states.  For our model, however,
the correlators of interest are strictly unaffected.
Formally, this follows since the transformations
\begin{eqnarray*}
  f(x) & \rightarrow & \exp\left[i\tau^z \mu_{\rm FM} 
x/v\right] f(x)\;, \\
  \psi(x) & \rightarrow & \exp\left[i \tau^z\left(\mu_{\rm QW} +
    \vec{h}\cdot{\vec{\sigma} \over 2}\right){x \over v}\right] \psi(x) \;,
\end{eqnarray*}
transform $\tilde{L}_{\rm FM} \rightarrow L_{\rm FM}$ and
$\tilde{L}_{\rm QW} \rightarrow L_{\rm QW}$, 
leave $F$ and $\Psi$ invariant, and respect the
boundary conditions at $x=0$.  Note that calculating expectation
values using these transformed fields (governed by $L_{\rm FM}$ and
$L_{\rm QW}$) in a functional integral formalism naturally produces
correlators normal ordered with respect to the shifted fields.  This
captures correctly the physics of filling/emptying the shifted energy
eigenstates discussed above.

$ $From the above discussion, it is apparent that the real-time
correlator appearing in Eqs.~(\ref{Jformula}) and (\ref{Iformula})
can be calculated using the pure, unpolarized Lagrangians, $L_{\rm FM}$ and 
$L_{\rm QW}$ corresponding to Eqs.~(\ref{LFM}) and (\ref{Lwire}), respectively.
Their $SU(2)$ invariance therefore implies
\begin{equation}   \label{SU2}
  \left\langle
    \left[F_\alpha^\dagger(0)\Psi_\beta^{\vphantom\dagger}(0),
      \Psi_\gamma^\dagger(t)F_\lambda^{\vphantom\dagger}(t) 
      \right]\right\rangle_{\tilde{H}_0} \Theta(-t) =
    \delta_{\alpha\lambda} \delta_{\beta\gamma} iC(-t)\;,
\end{equation}
where 
\begin{equation} \label{retarded}
  iC(t) = \Theta(t)\left\langle \left[
      B^{\vphantom\dagger}(t),B^\dagger(0) \right]\right\rangle
\end{equation}
is the standard retarded Green's function of the operator
\begin{equation}     \label{Bdef}
  B = F_\uparrow^\dagger \Psi_\uparrow\;.
\end{equation}
The choice of spin components in Eq.~(\ref{Bdef}) is arbitrary.

Substituting Eq.~(\ref{SU2}) into Eqs.~(\ref{Jformula}) and (\ref{Iformula}),
and using $U^\dagger(-t) = U(t)$, we find 
\begin{eqnarray*}
  \langle \vec{J}_{\rm tun}\rangle & = & - {\rm
    Im}\,\int_{-\infty}^\infty \! dt\, 
 {\rm Tr}\,\left(U(t) W^\dagger W^{\vphantom\dagger} \vec\sigma\right) 
 C(t)\;, \\
\langle I \rangle & = & - 2 \,{\rm Im}\,\int_{-\infty}^\infty \! dt\,
 {\rm Tr}\,\left(U(t) W^\dagger W^{\vphantom\dagger}\right) 
 C(t) \;.
\end{eqnarray*}
It is helpful to express the  matrices appearing in these
expressions in terms of
projection operators,
\begin{eqnarray*}
  W^\dagger W^{\vphantom\dagger} & = & \sum_{s=\pm 1} |t_s|^2
  \hat{u}_s \;, \\
  U(t) & = & \sum_{s'=\pm 1} \exp\left[ i \left(V + {{hs'} \over
        2}\right)t \right] \hat{v}_{s'} \;, 
\end{eqnarray*}
with $\hat{v}_s \equiv (1+s\hat{h}\cdot\vec\sigma)/2$ defined
analogously to $\hat{u}_s$, see Eq.~(\ref{pop}). 
 Then it becomes straightforward to
compute the averages
\begin{eqnarray*}
  {\rm Tr}\, \left(\hat{u}_s \hat{v}_{s'} \right) & = & {1 \over
    2}\left( 1 + s s' \hat{m}\cdot\hat{h}\right)\;, \\
  {\rm Tr}\, \left(\hat{u}_s \hat{v}_{s'} \vec\sigma\right) & = & {1
    \over 2}\left( s\hat{m} + s'\hat{h} + i s s' 
\hat{m}\times\hat{h}\right) \;,
\end{eqnarray*}  
and hence the tunneling spin current is
\begin{eqnarray}
  \left\langle \vec{J}_{\rm tun} \right\rangle & = & -{G \over 2}\sum_{s}
  \bigg[ (P\hat{m}+s\hat{h}) {\rm Im}\; \tilde{C}(V+hs/2+i\delta)
  \nonumber \\ 
  & & - P s\hat{m}\times\hat{h}\;{\rm Re}\,
  \tilde{C}(V+hs/2+i\delta)\bigg]\;.\label{Jeq1} 
\end{eqnarray}
Similarly, the charge current is
\[
  \left\langle I \right\rangle  = - G\sum_{s} 
  (1+P s\hat{m}\cdot\hat{h}) \;{\rm Im} \,
  \tilde{C}(V+hs/2+i\delta) \;. 
\]
The quantities $G$ and $P$ were defined in Eq.~(\ref{junctpar}), and 
we use the Fourier convention
\begin{equation}
  \tilde{C}(\omega) = \int\! dt \; C(t) e^{i\omega t}\;.
\end{equation}
The terms involving ${\rm Im}\; \tilde{C}$ are not surprising, since
this is directly proportional to the spectral function of $B$, and
hence has a simple interpretation in terms of tunneling via Fermi's
golden rule.  For these terms, we can use well-known results discussed
below.  However, the terms involving the real part of $\tilde C$
correspond to exchange processes generated by tunneling,
see Refs.~\onlinecite{PRL} and \onlinecite{slonc}.  That tunneling indeed
causes effective exchange couplings (even in the absence
of a ``bare'' exchange coupling) follows already from the
simple RG equations (\ref{RGeqns}).
As the physical effects of exchange are included
via the boundary condition  (\ref{bcc})
with the $SO(3)$ rotation matrix ${\cal R}(\vartheta)$,
we drop the terms proportional to ${\rm Re}\, \tilde{C}$ in the
spin current (\ref{Jeq1}). 

After some algebra, we then obtain the
tunneling spin current as 
\begin{equation} \label{tunnel}
\langle \vec{J}_{\rm tun} \rangle 
 = -{1 \over 2} \sum_s (P\hat{m}+s\hat{h}) \, {\cal
  I}_\alpha(V+hs/2, T) \;,
\end{equation}
where we have defined the function
\begin{eqnarray} \label{ialpha}
{\cal I}_\alpha(U,T) &=& G \,{\rm Im}\tilde{C}(U+i\delta)
\\ \nonumber &=&  G k_B T (k_B T/D)^\alpha \sinh\left(\frac{eU}{2k_B T }
\right) \\ \nonumber &\times&
 \left|\Gamma \left(1+\frac{\alpha}{2} + i \frac{eU}{2\pi k_B T}
\right)\right|^2\;,
\end{eqnarray}
with the bandwidth $D$ and the LL end-tunneling exponent
$\alpha>0$. We note that in SWNTs $D \approx
1$~eV, while $\alpha\approx 1.1$ according to the experiments
reported in Ref.~\onlinecite{LL-tubes2}. 
In the  limit $eU\ll k_B T$, the function (\ref{ialpha}) becomes 
\[
{\cal I}_\alpha \simeq \frac{GU}{2} (k_B T/D)^\alpha 
\Gamma^2(1+\alpha/2)  \;,
\]
and in the opposite limit,
\[ 
{\cal I}_\alpha \simeq \frac{GU}{2} (eU/2\pi D)^\alpha \;.
\]

The LL magnetization away from the
contact can now be related to the spin chemical potential
using the LL spin susceptibility $\chi=1/(2\pi v)$,
\begin{equation}\label{magnetization}
 \vec{M}= \vec{J}_R+\vec{J}_L= \chi\vec{h} \;.
\end{equation}
Thereby we obtain the spin current $\vec{J}$ 
injected into the LL from the FM at any given contact for arbitrary
exchange coupling $\vartheta$.  Omitting the expectation
values for brevity, we find
\begin{equation}\label{spinc}
\vec{J} =  {1 \over {2\pi}}{\cal  S}  \vec{h} + (1-{\cal S})
\vec{J}_{\rm tun}\;, 
\end{equation}
where ${\cal S}=({\cal R}-1)/({\cal R}+1)$ is a real antisymmetric
matrix.  Similarly, the injected charge current is
\begin{equation}\label{curr}
I  = - \sum_{s} (1+ sP\hat{m}\cdot\hat{h})\,{\cal I}_\alpha( V+hs/2,T)\;.
\end{equation}
For the case of a low-conductance contact, the exchange angle is small,
$\vartheta\ll 1$, and Eq.~(\ref{spinc}) can be further simplified to
\begin{equation} \label{sppcc}
\vec{J} =   \frac{\vartheta }{4\pi}
 \vec{h}\times \hat{m} - \frac12 \sum_s [P\hat{m}
+ s\hat{h}]\, {\cal I}_\alpha(V+sh/2,T)  \;.
\end{equation}
$ $From Eq.~(\ref{sppcc}), $\vartheta/4\pi$ can recognized as acting as a
sort of dimensionless spin conductance -- proportional in fact to the
``spin mixing conductance'' of Ref.~\onlinecite{brataas}.  The relations
(\ref{curr}) and (\ref{sppcc}), together with the results in
Sec.~\ref{sec:bulk}, provide the basis for our subsequent discussion
of the setup in Fig.~\ref{fig0}.

\section{Hydrodynamic description of bulk properties}
\label{sec:bulk}

In the previous section, we discussed how to determine the charge and
spin currents in the neighborhood of a contact in terms of the local
charge and spin chemical potentials $\mu$ and $\vec{h}$ of a LL.  To
complete the formulation of the full transport problem, we need to
understand how to relate these quantities at different points 
 {\sl  within} the LL.  This is the subject of this section.
In doing so, we assume that the length of the system is always long
compared to some characteristic dephasing length
beyond which the behavior is incoherent and hence classical.  For a
LL, we expect the dephasing length is set simply by the thermal
length scale $v/k_B T$.  At low temperatures, this length is very
long, and we thus are interested in the constitutive laws governing
the LL on long length scales.  

On such long length scales, we expect a rather classical description
to apply both to charge and spin.  For charge, classical behavior
follows due to dephasing.  For spin, more care must be taken, due to
the non-commuting nature of the spin operators.  On long length
scales, however, the total spin $\hbar s$ within any region is large.  For
large $s \gg 1$, all three components of the spin can in fact be
specified with very good accuracy, the  uncertainty being of 
$O(1/s)$.  Thus the 
long-wavelength ``hydrodynamic'' equations will involve simultaneously 
all three components of the magnetization and the corresponding currents.

\subsection{Operator equations of motion}

To construct the hydrodynamic equations, 
consider first the operator
equations of motion for the spin currents. 
 These are obtained from the usual Heisenberg equations,
$\partial_t \vec{J}_{L/R}(x) = i[H,\vec{J}_{L/R}(x)]$, 
with the Sugawara form of the spin Hamiltonian.\cite{book}
Writing 
\[
H_{\rm QW}=H_0+H_{\rm bs}+ H_{\rm magn}\;,
\]
the spin part of the LL Hamiltonian $H_0$ is
\begin{equation} \label{sug}
  H_0 = {v\over 2} \int \! dx\,  :\vec{J}_R\cdot\vec{J}_R 
    + \vec{J}_L\cdot\vec{J}_L:  \;.
\end{equation}
The backscattering contribution has already been specified in Eq.~(\ref{pert}),
and we now also include an external magnetic field $\vec{B}$ acting on the QW.
For that purpose, assuming a static and homogeneous field, we add the term
\begin{equation} \label{Hmag}
H_{\rm magn} = - \mu_B \int dx \vec{B} \cdot \vec{M} \;,
\end{equation}
with the LL magnetization (\ref{magnetization}).  Here
we have absorbed the electron $g$-factor $g_e$ into a renormalized
Bohr magneton $\mu_B= g_e e\hbar/2mc$.  The equations of motion are
\begin{eqnarray} \label{eom1} 
  (\partial_t - v\partial_x) \vec{J}_L & = &  b v\vec{J}_R
  \times \vec{J}_L - bv \partial_x \vec{J}_R -\mu_B \vec{J}_{L}\times
\vec{B}\;,  \\      \label{eom2}
  (\partial_t + v\partial_x) \vec{J}_R & = &   b v \vec{J}_L
  \times \vec{J}_R + bv \partial_x \vec{J}_L -\mu_B \vec{J}_{R}\times 
\vec{B} \;. 
\end{eqnarray}
Here terms like $\vec{J}_L\times \vec{J}_L$ are absent 
by virtue of normal-ordering.  Their absence follows also
by comparing the equations of motions obtained from the Sugawara
form (\ref{sug}) to the strictly equivalent
equations of motion of the corresponding $SU(2)$ level $k=1$
Wess-Zumino-Witten action.\cite{book}  

Taking the sum of Eqs.~(\ref{eom1}) and (\ref{eom2}), we find
\begin{equation} \label{sum1}
\partial_t \vec{M} + \partial_x \vec{J} =  -\mu_B \vec{M}\times
\vec{B}\;,
\end{equation}
where 
\begin{equation} \label{c_ren}
  \vec{J} = (1+b)v (\vec{J}_R - \vec{J}_L)\;.
\end{equation}
Comparing Eq.~(\ref{c_ren}) to Eq.~(\ref{defsc}), we see that
backscattering leads to a renormalization of the spin current.  This
is a ``backflow'' effect, similar to those in Fermi liquid theory.
For $B=0$, Eq.~(\ref{sum1}) represents the standard spin continuity equation.
Of course, the magnetic field will then spoil spin current conservation.

Taking the difference of Eqs.~(\ref{eom1}) and (\ref{eom2}) gives
\begin{equation} \label{precession1}
  \partial_t \vec{J} + (1-b^2)v^2 \partial_x \vec{M} = b v \vec{M}
  \times \vec{J} - \mu_B \vec{J}\times \vec{B}   \;.
\end{equation}
For $B=0$, we have a conserved spin current
$\vec{J}$ but bulk precession of the magnetization around the fixed
spin current.

\subsection{Hydrodynamics}
\label{sec:hydr}

Note that Eqs.~(\ref{sum1}) and (\ref{precession1}) are operator
identities, not equations for the expectation values of these
quantities.  We will call these expectation values the {\sl
  classical}\ values.  As argued at the beginning of this section, to
describe the bulk physics on long length scales, the hydrodynamics
should be phrased in terms of equations of motion for these classical
variables.  In the absence of backscattering, $b=0$, Eqs.~(\ref{sum1})
and (\ref{precession1}) are both linear, and so taking their quantum
expectation immediately gives the correct hydrodynamic description for
the classical values.  At zero temperature in the linear-response
limit, moreover, the RG analysis demonstrating that
$b$ is marginally irrelevant implies that this $b=0$
hydrodynamics remains qualitatively correct.  In general, even in this
case there will also be finite renormalizations of physical
quantities, but for $b \ll 1$, these are expected to be small.  

At $T>0$, and possibly also in non-linear response at $T=0$, however,
the hydrodynamic equations are generally corrected by dissipative
terms.  Formally, these exist due to the fact that $\langle
\vec{M}\times\vec{J} \rangle \neq \langle \vec{M}\rangle \times\langle
\vec{J}\rangle$.  Physically, dissipative corrections to
Eq.~(\ref{precession1}) describe processes caused by the backscattering 
in which a non-zero spin current can decay.  That the backscattering
interaction mediates such processes can be seen by expressing it
in terms of Fermions,
\begin{eqnarray}
  H_{\rm bs} & = & -{bv\over 4} \int \! dx\, \big[ 2\psi_{R\uparrow}^\dagger
  \psi_{R\downarrow}^{\vphantom\dagger} \psi_{L\downarrow}^\dagger
  \psi_{L\uparrow}^{\vphantom\dagger} + (\uparrow \leftrightarrow
  \downarrow) \nonumber \\
  & &  + \; \psi_R^\dagger \sigma^z\psi_R^{\vphantom\dagger}\,
  \psi_L^\dagger \sigma^z \psi_L^{\vphantom\dagger} \big]\;. \label{updown}
\end{eqnarray}
Consider an initial state containing one right-moving Fermion with up
spin and one left-moving Fermion with down spin, thereby carrying a
net spin current $J^z>0$ but no magnetization.  Acting on this state,
$H_{\rm bs}$ flips the spins of right- and left-moving
Fermions, thereby conserving the magnetization but {\sl reversing} the
spin current $J^z \rightarrow - J^z$.  Through a combination of such
processes, it is natural to expect a finite lifetime for the decay of
an initial spin current.  

\begin{figure}
\epsfxsize=0.8\columnwidth
\epsffile{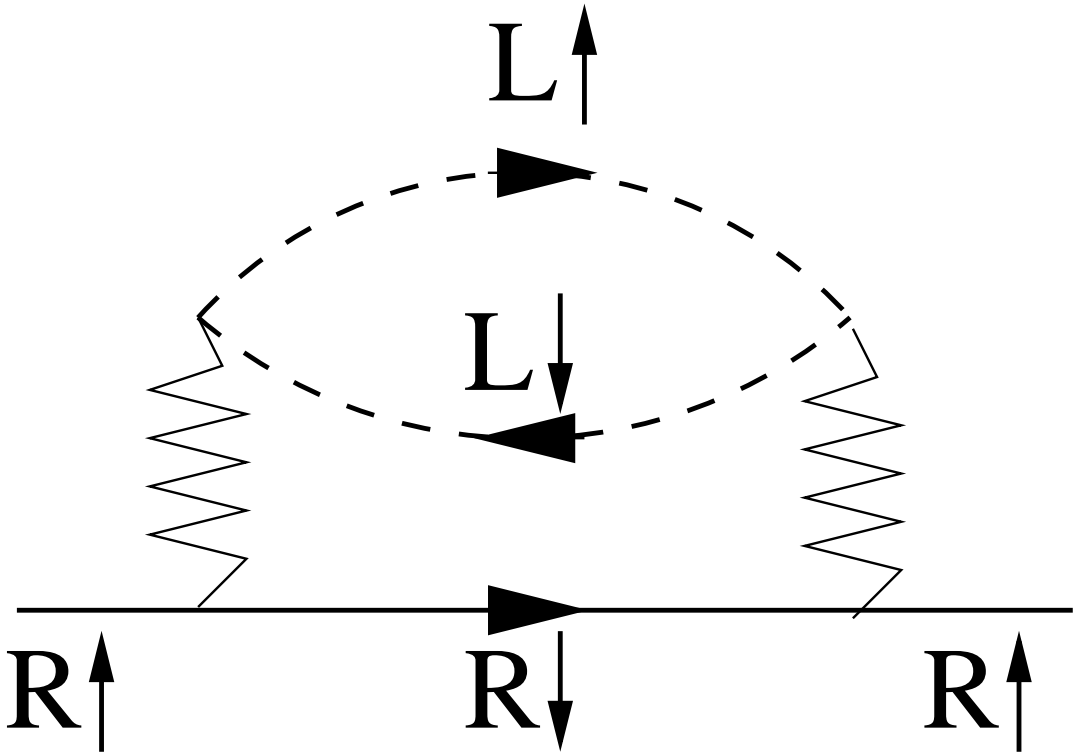}
\caption[]{\label{bubblefig} Self-energy diagram to estimate the decay 
  time for the spin current.}
\end{figure}

Further, on physical grounds, we can express the rate of such
decay processes based on Fermi's golden rule.  In particular, we
expect the inelastic spin current relaxation rate, $1/\tau^{\rm in}_J$,
to be proportional to an expectation value that is quadratic in 
$H_{\rm bs}$.  More formally, we can 
determine the lifetime using the leading non-constant correction 
to the Fermion self-energy, which is the usual two-loop bubble in
Fig.~\ref{bubblefig}.  Either way, since $1/\tau^{\rm in}_J$ is quadratic in
$b$, scaling determines the form of the decay rate,
\begin{equation} \label{curr_relax} 
  1/\tau^{\rm in}_J = A b^2 k_B T/\hbar \;,
\end{equation}
where we have used $k_B T$ to provide the energy scale needed from
scaling.  In addition, we have neglected more subtle logarithmic
corrections expected on general grounds.\cite{sachdev} The order unity
numerical prefactor $A$ is not obtained reliably by this simple
argument, but a crude estimate may be obtained using Fermi's golden
rule.  From the first term in Eq.~(\ref{updown}), we may consider the
rate for a single right-moving electron with down spin and momentum
$k$ to flip it's spin, simultaneously creating a left-moving
electron-hole pair:
\begin{eqnarray*}
  1/\tau_{\uparrow\downarrow} & = & \left( {bv \over 2}\right)^2 \int\!
  {{dk\, dq\, dq'} \over {(2\pi)^3}} (2\pi) \delta(vk-vk'-vq+vq')
 \\ & \times&
  (2\pi)\delta(k+k'-q-q') f(vk') \\
&\times& [1-f(vq)][1-f(vq')] \;,
\end{eqnarray*}
where $f(\epsilon)= 1/(e^{\epsilon/k_B T} + 1)$ is the Fermi
function.  This can be evaluated in the low-energy limit $vk \ll k_B
T$ to give the result in Eq.~(\ref{curr_relax}), with however a
surprisingly small prefactor $A=1/32\pi$.

We proceed at this stage on phenomenological grounds by modifying
Eq.~(\ref{precession1}) by hand to include spin current relaxation,
\begin{equation} \label{precession2}
  \partial_t \vec{J} + (1-b^2)v^2 \partial_x \vec{M} = -
  \vec{J}/\tau^{}_J + b v \vec{M} 
\times \vec{J} - \mu_B \vec{J}\times \vec{B}   \;,
\end{equation}
which provides a detailed (but approximate)
description of the crossover between ballistic and diffusive spin 
transport at low temperatures.  In general, elastic impurity
scattering processes will also relax the spin current, where in
fact no spin-orbit interaction is required.
This can be included according to Matthiessen's rule,
\begin{equation}
  {1 \over \tau^{}_J} = {1 \over \tau_J^{\rm in}} + 
{1 \over \tau_J^{\rm el}}\;.
\end{equation}
Since the same impurity scattering processes relax both spin and
charge currents, we expect $1/\tau_J^{\rm el}$ to be comparable to the
elastic scattering rate deduced from charge transport.

Taken together, Eqs.~(\ref{sum1}) and (\ref{precession2}) provide a
starting point for the investigation of the spin hydrodynamics of a LL
in the presence of electron-electron backscattering.  A useful check
is that these equations correctly recover the Landau-Lifshitz
dynamics\cite{ll}\ of classical ferromagnets at non-zero
temperatures.  Considering for simplicity $B=0$, in the low-frequency,
linear response limit, the non-linearity in Eq.~(\ref{precession2}) is
small, and the time-derivative of $\vec{J}$ can also be neglected.
Then one can solve for $\vec{J}$ perturbatively in the non-linearity
to obtain to leading order,
\begin{equation}
  \vec{J} = -D_s \partial_x \vec{M} - b v \tau^{}_J D_s \vec{M}
  \times \partial_x \vec{M} \; , \label{Ficks}
\end{equation}
where the spin diffusion constant is
\begin{equation}\label{grds}
  D_s = (v')^2 \tau^{}_J = \frac{\hbar v^2}{A b^2 k_B T} \;.
\end{equation}
For SWNTs, $b\ll 1$, and it is appropriate to approximate 
$1-b^2 \approx 1$. In addition,
the golden-rule estimate is at best valid for
$b \ll 1$, and we therefore 
ignore the tiny renormalization $v\to v'=(1-b^2)v$.  Inserting
Eq.~(\ref{Ficks}) into Eq.~(\ref{sum1}) then indeed gives the usual
Landau-Lifshitz equation.\cite{ll}\

We are predominantly interested in steady-state
situations in which both $\vec{M}$ and $\vec{J}$ are
time-independent.  Using Eqs.~(\ref{magnetization}) and
(\ref{precession2}), we find 
\begin{equation} \label{precdiff}
  \partial_x \vec{h} + \frac{1}{\sigma_s}\vec{J} 
  = (b/v) \, \left(\vec{h} + {{2\pi\mu_B} \over b}\vec{B}\right) \times
  \vec{J} \;,  
\end{equation}
where the linear-response spin conductivity is given by the Einstein relation 
\begin{equation}\label{spincond}
  \sigma_s = \chi D_s =  v \tau^{}_J/2\pi =
  \frac{\hbar v}{2\pi A b^2 k_B T} \;.  
\end{equation}
Note that the spin conductivity has dimensions of length as expected
in 1D, and is essentially given by the mean-free-path for
decay of spin currents, $\ell^{}_J = v \tau^{}_J$.

\section{Applications}
\label{sec:app1}

As an application of our general formalism, we now consider transport
for a LL connected to two FM reservoirs with magnetization directions
$\hat{m}_1$ and $\hat{m}_2$ for applied voltage $V$, see
Fig.~\ref{fig0}.  The FM magnetization unit vectors $\hat{m}_{1,2}$
are tilted by the angle $0\leq\theta\leq \pi$, so that $\hat{m}_1\cdot
\hat{m}_2= \cos\theta$.  For simplicity, we assume identical
low-conductance contacts on both sides such that $\mu_{\rm QW}=0$ and
the exchange angle is small, $\vartheta\ll 1$.  Furthermore, we assume
$P\ll 1$ for algebraic simplicity, although our results actually hold
somewhat more generally.\cite{foot1}\ The latter condition is
probably fulfilled in any practical application and ensures that the
spin chemical potential is small, $h/V\ll 1$.  The validity of the
latter condition is then self-consistently checked below.  For $h/V\ll
1$, we can use the following expansion for Eq.~(\ref{ialpha}),
\begin{equation}\label{expansion}
{\cal I}_\alpha( (V-sh)/2 ,T) = I_\alpha (V,T) -sh G_\alpha(V,T) \;.
\end{equation}
Here the current (per spin channel) for non-magnetic leads is 
\begin{equation} \label{ia}
I_\alpha(V,T) = {\cal I}_\alpha(V/2,T) \;,
\end{equation} 
such that for parallel FM magnetizations, $\theta=0$, 
the charge current $I(0)=2I_\alpha$ results in the absence
of backscattering.
Furthermore, the respective conductance is
\begin{equation} \label{gdef}
G_\alpha(V,T) =  dI_\alpha/dV = (e^2/2\pi \hbar) g_\alpha \;. 
\end{equation} 
Similarly we define a dimensionless contact conductance $g$
via $G=(e^2/2\pi \hbar) g$.

$ $From Eq.~(\ref{sppcc}), we can then write down the spin current
$\vec{J}_1$ through the left contact, taken at $x=0$ with the
local spin chemical potential $\vec{h}_1=\vec{h}(0)$,
and likewise $\vec{J}_2$ through the right  contact at 
$x=L$ with $\vec{h}_2=\vec{h}(L)$,
\begin{eqnarray} \label{j1}
\vec{J}_1 &=&  \frac{\vartheta}{4\pi}
 \vec{h}_1\times \hat{m}_1 + PI_\alpha \hat{m}_1 
-G_\alpha \vec{h}_1 \;,\\ \label{j2} 
\vec{J}_2 &=&  -\frac{\vartheta}{4\pi}
 \vec{h}_2\times \hat{m}_2 + PI_\alpha \hat{m}_2
+G_\alpha \vec{h}_2 \;.
\end{eqnarray}
The signs are chosen such that currents are oriented from left
to right.  Similarly, from Eq.~(\ref{curr}), 
the charge current flowing through the device follows,
\begin{equation} \label{current}
\frac{I(\theta)}{2I_\alpha} = 1 - \frac{P G_\alpha}{ I_\alpha}
 \vec{h}_1\cdot \hat{m}_1 \;,
\end{equation}
where we have exploited current conservation,
\begin{equation}\label{currcons}
\vec{h}_1\cdot\hat{m}_1 + \vec{h}_2\cdot \hat{m}_2=0 \;.
\end{equation}

Next these relations describing the
spin chemical potential and the spin current at the
boundaries need to be related by virtue of the hydrodynamic description
of Sec.~\ref{sec:bulk}.  
In the steady state, the basic hydrodynamic equations are
Eq.~(\ref{precdiff}) and 
\begin{equation} \label{sum}
\partial_x \vec{J}  =  -\frac{\mu_B}{2\pi v} \vec{h}\times \vec{B} \;,
\end{equation}
see Eq.~(\ref{sum1}).  We first consider transport in zero magnetic field,
and later on extend the analysis to finite fields in Sec.~\ref{sec:bb}.
In Secs.~\ref{sec:aa} and \ref{sec:bb}, the effects of
the backscattering interaction are neglected, $b\to 0$, so
that the spin resistivity vanishes.
This is expected to be appropriate for short-to-intermediate length $L$. 
The effects of $b\neq 0$ will then be addressed 
in Secs.~\ref{sec:cc} (without dissipation) and \ref{sec:dd}
(including spin diffusion).

\subsection{Effects of Spin-Charge Separation}
\label{sec:aa}

We start with the simplest case of 
$b=0$ and $B=0$, where the steady-state equations  
(\ref{precdiff}) and (\ref{sum}) are solved by a constant 
magnetization and hence spin chemical potential, $\vec{h}_1=
\vec{h}_2=\vec{h}$,
and conserved spin current $\vec{J}$.
To compute the charge current (\ref{current}),
we then need to find $\vec{h}$ which in turn is determined from
spin current conservation, $\vec{J}_1 = \vec{J}_2$.
Using Eqs.~(\ref{j1}) and (\ref{j2}), we then obtain three
equations, namely Eq.~(\ref{currcons}) and 
\begin{eqnarray*}
\frac{\vartheta}{4\pi}\vec{h} 
\cdot (\hat{m}_1\times\hat{m}_2) + G_\alpha\vec{h}
\cdot (\hat{m}_1-\hat{m}_2) && \\ - 2P  I_\alpha \sin^2(\theta/2)& =&0 \;,\\
\frac{\vartheta}{4\pi} \cos^2(\theta/2) 
\vec{h} \cdot (\hat{m}_1-\hat{m}_2) -
G_\alpha\vec{h}\cdot (\hat{m}_1\times\hat{m}_2) &=&0 \;.
\end{eqnarray*}
$ $From these relations, with $I(0)=2I_\alpha$,
the current results in the form
\begin{equation}\label{res1}
 \frac{I(\theta)}{I(0)}= 1- P^2 \frac{\tan^2(\theta/2)}{\tan^2(\theta/2)
 + Y_\alpha} \;.
\end{equation}
Here the quantity $Y_\alpha$ reads
\begin{equation} \label{yps}
Y_\alpha(V,T) =  1+ \left( \frac{\vartheta}
{2 g_\alpha}\right)^2 \;.
\end{equation}
For $eV\ll k_B T \ll D$, using Eq.~(\ref{gdef}), this becomes
\[
Y_\alpha\simeq \frac{(2\vartheta/g)^2}
{\Gamma^4(1+\alpha/2)} (k_B T/D)^{-2\alpha}  \;,
\]
while for $k_B T \ll e V \ll D$,
\[
Y_\alpha \simeq \frac{(2\vartheta/g)^2}
{(1+\alpha)^2} (e V/4\pi D)^{-2\alpha} \;.
\]
For a Fermi liquid, $Y_0 = 1 + (2\vartheta/g)^2$
is related to the dimensionless spin mixing conductance
$\eta$ of Ref.~\onlinecite{brataas}
by $Y_0= |\eta|^2/{\rm Re}(\eta)$, and Eq.~(\ref{res1})
correctly recovers the current-voltage relation 
of a Fermi liquid\cite{brataas} in the limit $\alpha\to 0$.

In the interacting case, $\alpha>0$, however,
for $eV, k_B T \ll D$, Eq.~(\ref{res1}) describes
a drastically different behavior.
Since $Y_\alpha$ becomes very large for low energies,
the {\sl spin accumulation effect is completely
destroyed}\ for any $\theta\neq \pi$. 
This remarkable phenomenon is entirely due to spin-charge separation,
since only then the exchange coupling is
so efficient at relaxing the injected polarized tunneling spin current.
The absence of spin accumulation is then a direct signature of 
the presence of spin-charge separation and would allow
to experimentally establish this phenomenon 
in a spin transport experiment.\cite{PRL}
Only for $\theta=\pi$, one gets the standard $1-P^2$ 
suppression in the current. 
At low applied voltage, $eV\ll k_B T$, the jump in $I(\theta)/I(0)$ 
from unity for $\theta<\pi$ down to $1-P^2$ at 
$\theta=\pi$ is only smeared out  
by thermal fluctuations, see Eq.~(\ref{yps}),
and therefore becomes very sharp at low temperatures.

Next we self-consistently check on the magnitude of $h/V$.
Multiplying $\vec{J}_1-\vec{J}_2=0$ by $\vec{h}$ yields 
with Eqs.~(\ref{j1}) and (\ref{j2}),
\begin{equation}
\frac{h}{V}=\frac{PI_\alpha}{2 V G_\alpha} \left( \frac{\tan^2(\theta/2)}
{\tan^2(\theta/2)+ Y_\alpha} \right)^{1/2} \;. \label{hmag}
\end{equation}
Note that $M=\chi h = h/(2\pi v)$, so that Eq.~(\ref{hmag}) also
describes the {\sl spin accumulation} in the QW. 
Now $I_\alpha/V G_\alpha$ equals $1$ at high temperatures,
and $1/(1+\alpha)$ for $k_B T\ll eV$.  Therefore $h/V \leq P/2$,
and for $P\ll 1$,  the assumed smallness of $h/V$ is
self-consistently verified.\cite{foot1}
Finally, we explicitly write down the spin current,
\begin{equation}
\vec{J} = \frac{PI_\alpha}{4} (\hat{m}_1+ \hat{m}_2)  
 \left  \{ 1 + (Y_\alpha-1)
 \frac{\tan^2(\theta/2)}{\tan^2(\theta/2)+ Y_\alpha} \right\} \;.
\end{equation}
Note that from Eq.~(\ref{currcons}), this implies
that $\vec{h}$ and $\vec{J}$ are orthogonal.

\subsection{Magnetic field dependence}
\label{sec:bb}

Next we consider a different
and probably more feasible experimental setup
aimed at revealing spin-charge separation.
Instead of changing the angle $\theta$ between
the FM magnetizations, a simpler setup could
work with a fixed angle $\theta$ but 
employ the additional  magnetic field $\vec{B}=B \hat{B}$. 
We assume that the bulk FM magnetizations $\hat{m}_{1,2}$ are 
not affected by this magnetic field,
and consider $b=0$.  Under these conditions,
the steady-state hydrodynamics in Eqs.~(\ref{precdiff}) and (\ref{sum})
describes a precession of both right- and left-moving 
spin currents around $\vec{B}$.  
Therefore, when moving from the left to the right contact,
the magnetic-field dependent precession phase   
\begin{equation} \label{precnew}
\gamma = \mu_B B L/ \hbar v 
\end{equation}
accumulates, so that $\gamma/2\pi$ is essentially the ratio of the
Zeeman energy to the level spacing.
Since a field of 1 Tesla corresponds to
0.058~meV, a sizeable precession phase can
easily be achieved for tube lengths in the micron range and 
standard magnetic field strengths. 

$ $From Eqs.~(\ref{precdiff}) and
(\ref{sum}), we can relate $\vec{h}_2$ and $\vec{J}_2$
at the right contact to the respective
quantities at the left contact.  Some algebra leads to
\begin{equation} \label{hh2}
\vec{h}_2= \cos\gamma\, \vec{h}_1 + (1-\cos\gamma) \,
(\hat{B}\cdot \vec{h}_1)\, \hat{B} - \sin\gamma \, \vec{J}_1\times \hat{B}\;,
\end{equation}
and similarly 
\begin{equation}\label{jj2}
\vec{J}_2= \cos\gamma \, \vec{J}_1 + (1-\cos\gamma) \,
(\hat{B}\cdot \vec{J}_1)\, \hat{B} - \sin\gamma \, \vec{h}_1\times \hat{B}\;.
\end{equation}
To compute the current, we then have to compute
$\vec{h}_1\cdot \hat{m}_1$ as outlined in Appendix \ref{appendixa}.
This is in general cumbersome and requires a numerical
analysis.  We thus restrict our attention to the
special case $\theta=\pi$, where the 
FM magnetizations $\hat{m}_1=-\hat{m}_2$ are 
antiparallel.  
Note that for $\theta<\pi$, the spin accumulation effect is quenched
anyways by spin-charge separation, and then the magnetic field should have
virtually no effect for $\alpha>0$.  The case $\theta=\pi$ is therefore
also of most interest.

First we consider a {\sl fixed field strength}\ $B$.
Let us suppose that $B$ has been adjusted such that the precession phase is
$\gamma=\pi$, but one allows for an arbitrary field 
direction $\hat{B}$.  In that case,
the equation in Appendix~\ref{appendixa}
can be solved analytically, and the current is
\begin{equation}\label{beta}
 I(\beta)/2I_\alpha =  1- P^2  \frac{1}{1 + Y_\alpha \tan^2\beta}  \;,
\end{equation}
where $Y_\alpha(V,T)$ is defined in Eq.~(\ref{yps}) and
$\hat{B}\cdot\hat{m}_1 =\cos\beta$.
If the magnetic field is parallel to the FM magnetizations,
$\beta \to 0$, a maximum spin accumulation
effect is recovered, $I(\beta)/2 I_\alpha=1-P^2$.
Upon tilting the magnetic field, however, since the quantity $Y_\alpha$
diverges at low energies,  spin accumulation
is quenched again.
This is very similar in effect to varying the
angle $\theta$ between the FM magnetizations, 
but may be easier to implement experimentally.
For $\beta = 0$, there is no precession since $\vec{h}_1$ is parallel
to $\hat{m}_{1,2}$ and hence to $\hat{B}$. Therefore spin accumulation
is not affected by the magnetic field.  $ $From this observation,
we then expect that the suppression
of spin accumulation at $\beta\neq 0$ holds in fact for 
a broad regime of magnetic field strengths, and no special fine-tuning 
to $\gamma=\pi$ should be necessary.
The behavior of $I(\beta)$ is {\sl qualitatively}\
different depending on whether spin-charge separation is realized
or not.

Another possibility consists of fixing the {\sl field direction} $\hat{B}$
and then measuring the $I-V$ characteristics for different magnetic
field strength $B$ or,
equivalently, precession phase $\gamma$, see Eq.~(\ref{precnew}).
For simplicity, let us assume that $\hat{B}$ is adjusted
perpendicular to $\hat{m}_1$.  Under this condition, 
the current can be found in closed form again, with the result
\begin{equation}\label{currbb}
\frac{I(\gamma)}{2I_\alpha} = 1 - P^2 \frac{1}{1+F(\gamma)}\;,
\end{equation}
where we use 
\begin{equation}\label{kk}
F(\gamma)= \frac{(2\pi/g_\alpha)^2 -1+Y_\alpha 
\cos^{-2}(\gamma/2)}{(g_\alpha/2\pi)^2 Y_\alpha+ \cot^2(\gamma/2)} \;.
\end{equation}
Note that Eq.~(\ref{currbb}) predicts a sharp {\sl negative transverse 
  magnetoresistance} peak near $B=0$ (and again at periodic
intervals).  In particular, at low temperatures, and in the linear
bias regime, $Y_\alpha, g_\alpha^{-1} \gg 1$, and one can estimate the 
full-width half-maximum (FWHM) of this feature in $I(B)$ by
\begin{equation}
  \left(\Delta \gamma\right)_{\rm FWHM} = {2g_\alpha \over \pi}
  \left[1 + \left( 
      {\vartheta \over {4\pi}}\right)^2 \right]^{-1/2},
\end{equation}
which therefore decreases as $T^\alpha$ as $T \rightarrow 0$.  The
{\sl height} of the peak is, however, temperature independent.  More
generally, the current (\ref{currbb}) is a periodic function of the
precession phase $\gamma$, with $F(\gamma)=0$ for $\gamma$ equal to an
integer multiple of $2\pi$.  In that case, the spin accumulation
effect is maximal, and thus the current minimal,
$I(\gamma)/2I_\alpha=1-P^2$.  On the other hand, for $\gamma$ being an
odd multiple of $\pi$, $F(\gamma)\to \infty$ and thus a complete
suppression of spin accumulation obtains.

\begin{figure}
\epsfxsize=0.8\columnwidth
\epsffile{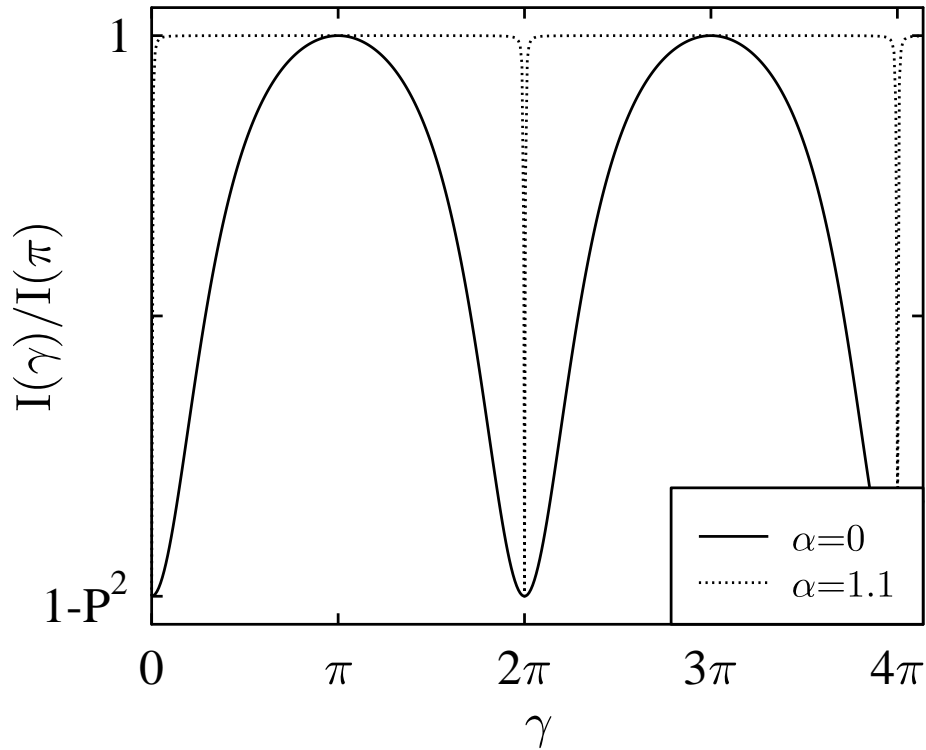}
\caption[]{\label{fig1} 
Magnetic field dependence of the $T=0$ current for 
$\theta=\pi$, $G= 0.08 e^2/h$, $eV/D=0.1$, $\vartheta=0.5$, both
for a Fermi liquid ($\alpha=0$) and for a LL with $\alpha=1.1$,
where $\vec{B}\perp \hat{m}_1$. 
The precession phase $\gamma\sim B$ is given 
in Eq.~(\ref{precnew}).  }
\end{figure}

Remarkably, the way that one interpolates between these
two limiting cases by varying $\gamma$ 
strongly depends on the correlation strength
$\alpha$, see Fig.~\ref{fig1}.
The $\gamma$ dependence of the current-voltage relation
can therefore  again reveal spin-charge separation. 
While for $\alpha=0$, the current is a smooth periodic function
in $\gamma$, for $\alpha>0$ and low energies, the periodicity is
turned into a series of very sharp dips at $\gamma$ equal to integer
multiples of $2\pi$.  This can be seen from Eq.~(\ref{kk}), yielding
for $T=0$ and $eV\ll D$,
\[
F(\gamma) \sim V^{-4\alpha}
 \frac{1+(\vartheta/4\pi)^2 
\cos^{-2}(\gamma/2)}{(\vartheta/4\pi)^2+\cot^2(\gamma/2)} \;.
\]
Therefore the spin accumulation effect is quenched 
unless $\cot(\gamma/2)$ diverges.  
Measuring the magnetic field difference corresponding to
the distance $\Delta \gamma=2\pi$ between two dips 
can also provide a direct estimate of the contacted length $L$ 
from Eq.~(\ref{precnew}).

\subsection{Backscattering: Dissipationless Precession}
\label{sec:cc}

For the remainder of this section, we address
the consequences of a finite backscattering coupling
$b$, taking for simplicity $B=0$.  We start with the
zero-temperature limit where the spin resistivity vanishes and
Eq.~(\ref{precdiff}) predicts  bulk precession of $\hat{h}(x)$
around the fixed spin current,
\begin{equation} \label{precession}
\partial_x \hat{h} = (b/v) \, \hat{h}\times \vec{J} \;.
\end{equation}
Clearly, $\hat{h}\cdot \hat{J}$ is also conserved.
It is then useful to introduce the quantities
\begin{eqnarray} \label{delta}
 C_b &=& \frac{b g P  L D}{4 \hbar v} \;,
\\ \label{gamma}
 C_\vartheta  &=& \left( \frac{2\vartheta}
{(1+\alpha) g} \right)^2 \;,
\end{eqnarray}
which provide dimensionless
measures of the backscattering
strength $(C_b)$ and of exchange ($C_\vartheta$).
The problem is then fully specified in terms of four 
dimensionless parameters, namely the LL exponent
 $\alpha$, the tilting angle $\theta$, and of course
$C_b$ and $C_\vartheta$.

Similar to Sec.~\ref{sec:bb}, precession of $\hat{h}(x)$ 
around the conserved spin current implies a precession phase
when going from the left to the right contact,
\begin{equation} \label{phase}
\Delta \varphi = b J L/ v \;.
\end{equation}
The spin chemical potentials $\vec{h}_1$ and $\vec{h}_2$
can now be determined from spin conservation, 
$\vec{J}_1=\vec{J}_2$,
and the precession equation (\ref{precession}).
In addition, a self-consistency condition
arises since $J$ appears itself in the precession 
phase (\ref{phase}).
This self-consistency implies that we are dealing
with a {\sl nonlinear} transport problem 
for $b>0$ and has far-reaching consequences.
For convenience, use the abbreviation
\begin{equation} \label{abb1}
X= \frac{hG_\alpha}{PI_\alpha} \;,
\end{equation}
with $0\leq X\leq 1$, which
provides a dimensionless measure of the absolute value of 
the spin chemical potential. Then
the charge current (\ref{current}) is
\begin{equation}\label{curr0}
I= 2I_\alpha (1- P^2 X^2) \;.
\end{equation}
$ $From Eq.~(\ref{j1}), the absolute value of the spin current 
can be expressed in terms of $X$ as well,
\begin{equation}\label{jabs}
J^2  = (PI_\alpha)^2 \left[1+\left(\frac{\vartheta X}
{2g_\alpha}\right)^2\right] (1-X^2) \;,
\end{equation}
where $g_\alpha$ is given in Eq.~(\ref{gdef}).
In Appendix~\ref{appendixb}, we derive the following
self-consistency equation determining the current,
\begin{equation}\label{scs1}
\sin^2(\theta/2) =  \frac{X^2
+(\vartheta X/2 g_\alpha)^2}{1+
(\vartheta X/2g_\alpha)^2}
\cos^2(\Delta\varphi/2) \;.
\end{equation}
Using the precession phase (\ref{phase}) 
with Eq.~(\ref{jabs}), one can then solve
for $X$ and thereby obtain the $I-V$ relation
(\ref{curr0}).

Remarkably, solution of Eq.~(\ref{scs1}) predicts  a 
{\sl multi-valued}\ $I-V$ relation.  This is a direct
consequence of the non-linearity of this spin transport
problem in the presence of backscattering.
Under an exact calculation, we would 
in fact expect a unique answer for the
current, since thermal or quantal fluctuations around 
Eq.~(\ref{scs1}) should stabilize just one solution  
out of the multiple branches.
Such a calculation could be performed
in principle following a Keldysh approach or employing
Langevin-type equations, but seems difficult to pursue in practice.
A simpler approach based on an (approximate)
free energy principle could obtain from the analogy 
to SNS junctions in Sec.~\ref{sec:sns} or the
 related ``tilted washboard'' picture
of Josephson junctions above the  critical current.
 In the latter system, 
a free-energy principle is well-known to apply and to
correctly resolve multistability questions posed
by the equations of motion alone.\cite{likharev} 
The difficulty with this approach for the spin transport
problem at hand is to describe the dissipative terms in
such a free energy.  While the bulk energy is known,
the boundary terms are less straightforward to handle,
and, unfortunately, a rigorous resolution of 
this question has to  remain open.  Below we shall
argue on intuitive grounds as to which of the multiple
branches is realized.

In general, one needs to numerically find the solutions 
to Eq.~(\ref{scs1}), e.g.~using a Newton-Raphson root-finding 
algorithm.\cite{recipes} 
Numerical results accurately confirm an analytical
solution possible for low energies, $eV\ll D$, and $\alpha>0$.
In the following, we focus on this regime of most interest,  
where $Y_\alpha\gg 1$, see Eq.~(\ref{yps}),
and search for solutions that fulfill also
$Y_\alpha X^2 \gg 1$.  Note that otherwise the effect 
of backscattering is negligible in any case.
Under these conditions, the self-consistency equation 
(\ref{scs1}) is solved by the precession phase taking only
one of the following discrete values,
\begin{equation} \label{special}
\Delta \varphi= (2n+1)\pi - \theta\;,
\end{equation}
where $n=0,1,2,\ldots$ is a {\sl winding number} counting 
the number of full precession cycles of the steady-state
bulk magnetization as one proceeds
from the left to the right contact. 
In principle, there is also a set
of solutions obtained from the substitution $\theta\to -\theta$
in Eq.~(\ref{special}).  For $\theta=0$ and $\theta=\pi$,
both sets coincide, and for $0<\theta<\pi$ we expect that
only Eq.~(\ref{special}) gives stable solutions.

Then the following general picture emerges.
Focussing for concreteness on the case $\theta=0$,
the self-consistency equation (\ref{scs1}) is
solved either by $X=0$ or by $\Delta\varphi=(2n+1)\pi$.
As the voltage $V$ is increased, first we have an
arbitrary precession phase $\Delta \varphi<\pi$ that
increases with $V$.  At the same time, Eq.~(\ref{scs1})
enforces $X=0$, leading to the standard
$b=0$ current $I=2I_\alpha$.  As the precession phase
hits $\Delta \varphi=\pi$ at the voltage $V=V_0(\theta)$ 
(see below), the spin current $J$ is locked at a fixed
value such that $\Delta \varphi$ remains constant 
when further increasing the voltage.  To keep $J$
constant, however, the charge current $I$ (or, 
equivalently, the quantity $X$) has to adjust
from Eq.~(\ref{jabs}).  Since this leads to
a quadratic equation for $I$, there are two
possible solutions for $I$. However, one of them 
would lead to unphysical currents exceeding
$2I_\alpha$, and is disregarded in what follows.
As the voltage is now increased up to $V_1(\theta)$, 
the precession phase $\Delta \varphi=3\pi$
becomes possible, and the above picture is
re-iterated.

For arbitrary $\theta$, Eq.~(\ref{special}) then predicts 
the following current-voltage relation.
For $V<V_0(\theta)$, the $b=0$ current (\ref{res1}) is realized.
Upon increasing the voltage above the
threshold $V_0(\theta)$, however, {\sl sawtooth-like oscillations}
appear.  In the window $V_n(\theta)<V<V_{n+1}(\theta)$, we obtain
\begin{eqnarray} \label{spec2}
\frac{I(V)}{2I_\alpha} &=& 1 - \frac{P^2}{2} \Biggl\{1- 
\frac{(eV/4\pi D)^{2\alpha}}{C_\vartheta} \\ \nonumber &+& 
\Biggl [ \left(1+\frac{(eV/4 \pi D)^{2\alpha} }{C_\vartheta}
\right)^2 \\ \nonumber &-& \frac{1}{C_\vartheta}
 \left(\frac{(2n+1)\pi-\theta}{C_b (eV/4\pi D)}\right)^2 
\Biggr]^{1/2} \Biggr\}\;.
\end{eqnarray}
In the low-energy limit, the voltages $V_n(\theta)$
are given by
\begin{equation}   \label{tv}
\frac{e V_n(\theta)}{4\pi D}= \frac{(2n+1)\pi-\theta}
{C_b \sqrt{C_\vartheta}} \;,
\end{equation}
and the $I-V$ relation for $V_n(\theta)
< V< V_{n+1}(\theta)$ simplifies to
\begin{equation}\label{osciv}
\frac{I(V)}{2I_\alpha}= 1- \frac{P^2}{2} \left(1 +
\sqrt{1-(V_n(\theta)/V)^2} \right) \;.
\end{equation}
Note that the $V_n(\theta)\sim 1/L$ and hence 
can in principle be made arbitrarily small simply
by increasing the QW length $L$. 
Therefore the effects of backscattering become very
important in sufficiently long QWs.  Estimating the period
$\Delta V=V_{n+1}-V_n$ corresponding to a full precession
cycle from Eq.~(\ref{tv}) for typical SWNT parameters, we find 
$\Delta V\approx 10$ to 100 mV.
Measuring the oscillation period $\Delta V$
could provide  useful information about the backscattering
interactions and the exchange angle.

\begin{figure}
\epsfxsize=0.8\columnwidth
\epsffile{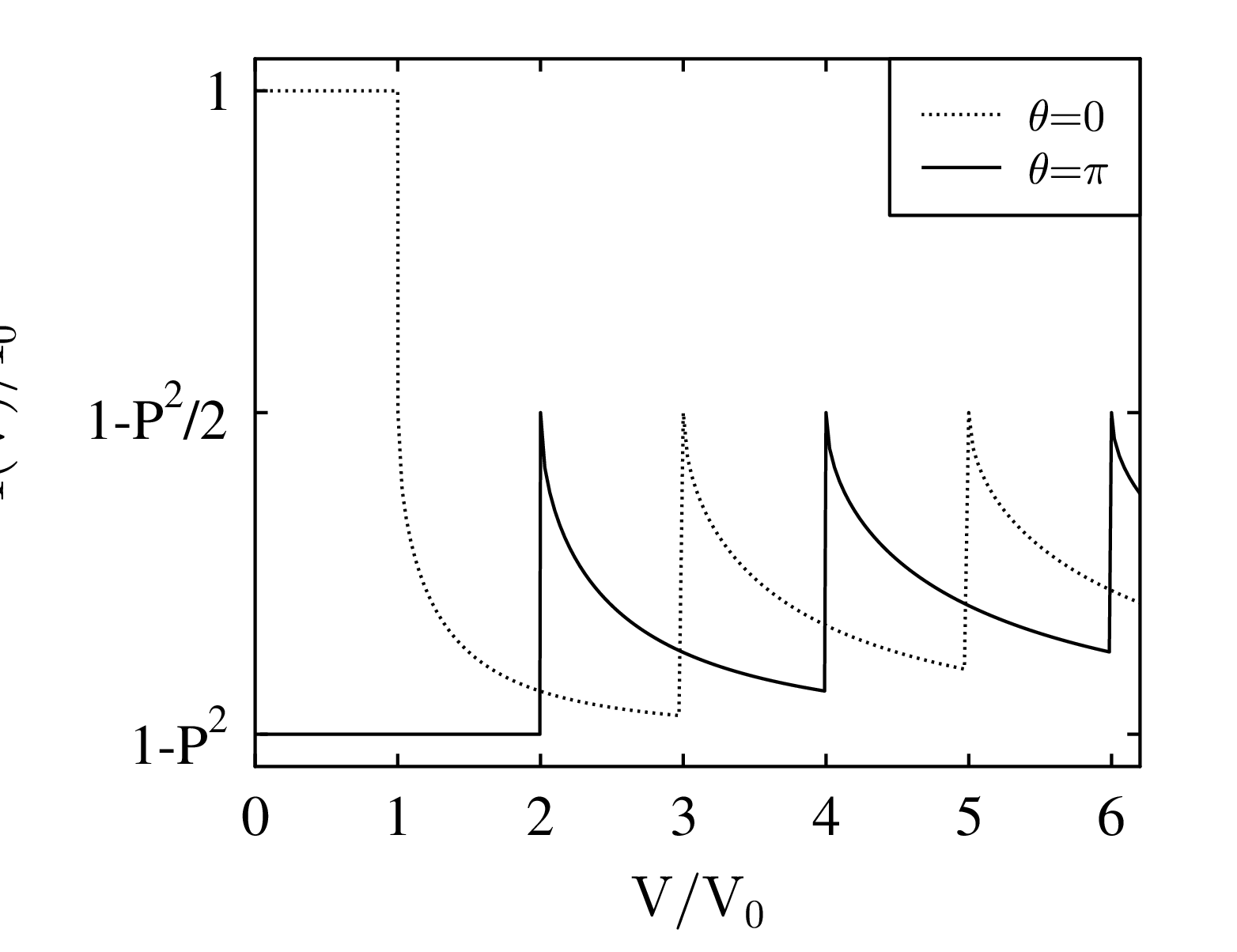}
\caption[]{\label{fig4} 
Current-voltage relation in the presence of
backscattering for $T=0$ and $\theta=0,\pi$. 
The scales are set by $I_0=2I_\alpha$ and  $V_0=V_0(\theta=0)$.  }
\end{figure}

The non-sinusoidal oscillatory $I-V$
relation (\ref{osciv}) is depicted in Fig.~\ref{fig4} for typical
SWNT parameters and $\theta=0, \pi$.  Apparently,
backscattering has a dramatic influence
on spin transport not anticipated from
thermodynamical considerations. 
To experimentally observe the 
predicted sawtooth-like oscillatory $I-V$ relation, 
however, it will probably be necessary to 
measure at very low temperatures 
using rather long and clean QWs.
Note that these oscillatory behaviors are a 
non-equilibrium effect not present in the temperature
dependence of the linear conductance.

\subsection{Finite-temperature dynamics}
\label{sec:dd}

As outlined in Sec.~\ref{sec:hydr}, for finite temperature,
backscattering causes spin diffusion, and we now
address the effects of spin diffusion on the current-voltage
relation, focussing on zero magnetic field, $B=0$, 
where the spin is still conserved.
The self-consistency equation in this case is derived
in detail in Appendix~\ref{appendixc}.
To solve this lengthy equation, we again restrict ourselves
to the low-energy regime, $k_B T,  eV \ll D$, with $\alpha>0$.

For $T\ll T^*$, we then basically recover the results of Sec.~\ref{sec:cc},
while for $T\gg T^*$, with the crossover temperature 
\begin{equation} \label{tstar}
T^* = \frac{\hbar v}{2\pi^2 k_B A b^2 L \vartheta} \sim 1/L \;,
\end{equation}
a spin-diffusion-dominated regime emerges. 
This temperature is defined by $\pi \vartheta L = \sigma_s(T^*)$,
see Appendix~\ref{appendixc}, with the spin conductivity (\ref{spincond}).
In the following, we focus on the regime $T\gg T^*$,
for which the $I-V$ relation can be written as
\begin{equation} \label{ivnew}
\frac{I(V)}{2I_\alpha} = 1- \frac{P^2}{2}
 \left(1+ \sqrt{1- A_n} \right) \;,
\end{equation}
where 
\begin{equation}
A_n=  \frac{2 \hbar G_\alpha } {\sigma_s L}
\left( \frac{[(2n+1)\pi - \theta] e v}{b PI_\alpha} \right)^2\;.
\end{equation} 
Remarkably, since $\sigma_s\sim 1/b^2$,
any dependence on the backscattering coupling $b$ drops 
out in this temperature regime.
In principle, the current is then again multivalued and indexed
by a winding number $n$.  
Here the appropriate threshold voltages, above which the respective
current can be realized, follow from the condition
$A_n\leq 1$.  On physical grounds, in this spin-diffusion limited
transport regime, we 
expect that only the lowest winding number $n=0$ is realized,
leading for $V>V_0(\theta)$ to the current,
\begin{equation}
\frac{I(V)}{2I_\alpha} = 1- \frac{P^2}{2} 
 \left(1 + \sqrt{1- (V_0(\theta) /V)^{2+\alpha}} \right) \;,
\end{equation}
where we assume $k_B T\ll eV$ and use the voltage scale
\begin{equation}
\left(\frac{eV_0 (\theta)}{4\pi D}\right)^{2+\alpha} =  \frac{(1+\alpha) 
\hbar e^2 v^2 (1- \theta/\pi)^2}{2\sigma_s L b^2 P^2 G D^2}
\;. \label{uglymess} 
\end{equation}
Note that the right-hand-side of Eq.~(\ref{uglymess}) is  proportional
to $(k_B T/D) ( \hbar v/LD)$, with a prefactor of order unity.
Hence this spin-diffusion limited regime should be accessible
to experiments.

\section{Extensions}
\label{sec:extensions}

The approach developed in this paper is very flexible, and can
straightforwardly be extended to describe a variety of other
physically relevant situations.  We sketch some of these extensions in
this section.

\subsection{Bulk contacts}

In many cases, particular in experiments on carbon nanotubes, contacts
are made not to the ends of the quantum wire but to points in the
``bulk''.  Much of the preceding theory applies to this case as well,
but there are some additional complications.  First, let us reconsider
the problem of a single contact at $x=x_1$, this time in the bulk
($0<x_1<L$) rather than at the boundary.  Because there is no boundary
condition relating right- and left-moving fields, there are more
independent couplings.  In
general, the contact Hamiltonian $H_c$ has three distinct
contributions, neglecting redundant forward scattering terms which
give spin-independent phase shifts, 
\[
H_c = H_{\rm tun} + H_{\rm ex} + H_{\rm bs1} \;.
\]
Tunneling is described by the tunneling Hamiltonian (\ref{tun}),
where $\Psi = \psi(x_1)$, and we allow for different hopping
matrix elements for right- and left-moving states of each spin
polarization, $t_s^{(a=R/L)}$.
There are also two distinct exchange terms,
\begin{equation}
  H_{\rm ex} = - \sum_{a=R/L} K_a \,
\hat{m} \cdot \Psi_a^\dagger {{\vec\sigma}
    \over 2} \Psi_a^{\vphantom\dagger}\;.
\end{equation}
Finally, there are local single-particle backscattering processes
described by
\begin{equation}
  H_{\rm bs1} = \sum_s \xi_s \Psi_R^\dagger \hat{u}_s\cdot\vec\sigma
  \Psi_L^{\vphantom\dagger} + {\rm h.c.}\ \;,
\end{equation}
where the projection operator $\hat{u}_s$ is given in Eq.~(\ref{pop}).
These terms arise since the presence of a contact inevitably 
leads to local disorder within the QW.

For a non-interacting QW, all three contributions are on an equal
footing, as they all involve Fermion bilinears.  When interactions are 
present in the QW, however, this is no longer the case.  In
fact, in the boundary RG framework, they scale completely
differently.  In particular, the tunneling Hamiltonian $H_{\rm tun}$ is 
  irrelevant, the boundary exchange $H_{\rm ex}$ is marginal, and
the single-particle backscattering $H_{\rm bs1}$ is  relevant.  
Thus provided all terms in $H_c$ are comparable, $H_{\rm bs1}$ will
dominate at low energies, where the distance to another
contact, the inverse temperature, and any inverse voltage are all
sufficiently large.  The effect of such relevant backscattering terms
has been studied extensively.\cite{kane92} We expect that the final
result in the low energy limit is to completely sever the LL
into two halves at $x=x_1$.  In that case, one can
effectively ignore at very low energies half of the LL, namely the one
  not connected to a closed circuit, and treat the other half using the
end-contact phenomenology of the previous sections.

The latter discussion presumes that the $\xi_s$ are substantial.  In many 
cases, however, it is natural to expect that in fact
$|\xi_s| \ll |K_a|,
|t_s^{(a)}|$.  For nanotubes, if the characteristic scale
of the contacts is substantially larger than the inter-atomic
dimensions of the nanotube, the $\xi_s$,
involving matrix elements that oscillate on the atomic scale, are
considerably suppressed.  Similarly, in contacts to semiconductor
quantum wires with widths large compared to the Fermi wavelength,
$\xi_s$ can be suppressed due to the smoothness of the effective
potential at the contact.  
It is therefore of interest to describe the problem in
the absence of single-particle backscattering, $\xi_s=0$.
The equation-of-motion methods of Secs.~\ref{sec:contacts}
 and \ref{sec:bulk} can then be
straightforwardly extended to describe bulk contacts.  For
illustration purposes, consider in particular the case of multiple
tunneling contacts at points $0<x_i<L$, neglecting for simplicity bulk 
backscattering.  Then the equations of motion for the chiral currents
are 
\begin{eqnarray} \label{bulkeom}
  (\partial_t + av \partial_x) \vec{J}_a & = & - \mu_B \vec{J}_a
  \times \vec{B} \\ \nonumber 
  &+& \sum_i \delta(x-x_i) \left[\vec{J}_a^{\rm tun}(x_i) + K_a
    \hat{m}_i \times \vec{J}_a\right]\;, 
\end{eqnarray}
where $a=L/R=\mp$, and the tunneling current $\vec{J}^{\rm
  tun}_a(x_i)$ is determined from Eq.~(\ref{tunnel}), with chemical 
potentials
$V$ and $\vec{h}$ appropriate to that contact and chirality.
Furthermore, in Eq.~(\ref{tunnel}), one has to replace the
end-tunneling exponent $\alpha$ by the bulk-tunneling exponent 
$\alpha_{\rm bulk}<\alpha$.\cite{book}
Similarly, the chiral charge
currents $I_{R/L}$ obey the equations of motion,
\begin{equation} \label{bulkeom2}
  (\partial_t + a v_c \partial_x)I_a = \sum_i \delta(x-x_i) I_a^{\rm
    tun}(x_i)\;, 
\end{equation}
where $v_c$ is the charge velocity.
Each pair of equations can be combined to give equations for the spin
and charge densities and currents, which can be solved in the steady
state knowing the tunneling spin and charge currents at each contact
(determined by the voltages) and the boundary conditions $I_R = I_L$
and $\vec{J}_R=\vec{J}_L$ at $x=0$ and $x=L$. 
Precession, diffusion, and spin-orbit scattering
(see below) can also be included simply by adding the appropriate
terms to the right-hand-sides of Eqs.~(\ref{bulkeom}) 
and (\ref{bulkeom2}).

\subsection{Nanotubes and Flavor}

While the general methodology and physical results applied above
pertain to any interacting QW, some differences exist that
should be taken into account in applying the formalism in detail to
nanotubes.  In particular, a ``pristine'' SWNT has not one but two 
1D bands crossing the Fermi energy, arising from the
sublattice reflection symmetry of the graphene lattice.  Thus in fact
the low-energy description of SWNTs requires an
additional ``flavor'' index $A=1,2$ on all electron fields,
 $\psi_{a\alpha} \rightarrow \psi_{aA\alpha}$.
 Moreover, the low-energy Hamiltonian
describing the nanotube in the absence of backscattering
and magnetic fields respects the full chiral $U(4) \times U(4)$ symmetry of
arbitrary separate $U(4)$  rotations of right- and left-movers in the
combined spin-flavor space.  This symmetry implies that, to a good
approximation (since backscattering terms are weak), not
only the $SU(2)$ spin currents $\vec{J}_{R/L}$ but the full $SU(4)$
{\sl spin-flavor currents},
\begin{equation}
  J^{A\alpha B\beta}_a = \;:
  \psi^\dagger_{aA\alpha}\psi^{\vphantom\dagger}_{aB\beta}   : \;,
\end{equation}
are conserved.  In the ballistic limit, these currents
satisfy chiral wave equations away from the contacts,
\begin{equation} \label{chiralSU4} 
  (\partial_t \mp v \partial_x) J^{A\alpha B\beta}_{a} = 0 \;,
\end{equation}
and a full solution of the transport problem in the steady state requires
imposing constant values of each of these ($16+16=32$) chiral
currents between contacts and/or ends of the nanotube.   Moreover,
$SU(4)$ generalizations of the contact exchange $H_{\rm ex}$ 
can be expected.

As we have seen in the simpler single-channel case above,
backscattering terms, even when weak, can lead to significant effects
in long tubes.  Because the backscattering interactions do not respect
the (accidental) $SU(4)$ symmetry but only the physical symmetries,
there are in fact a variety of independent backscattering couplings.
Thus, in general, the extension of $H_{\rm bs}$ to include flavor is
rather complicated -- the interested reader will find the ($8$ or
$11$, depending upon whether the nanotube is undoped or doped) diverse
backscattering terms enumerated in
Refs.~\onlinecite{LL-tubes1} and
 \onlinecite{balents96}.  These
interactions lead to a variety of generalized ``precession'' terms in
the operator equation of motion, which must be added to
Eq.~(\ref{chiralSU4}).  These have the general form
\begin{eqnarray*}
   (\partial_t \mp v \partial_x) J^{A\alpha B\beta}_{a} & = & 
   f_{BDEF}^{(a)} J_R ^{A\alpha D\nu} J_L^{E\beta F\nu}  \\ 
   & - & f_{CAEF}^{(a)} J_R^{C\mu B\beta} J_L^{E\mu F\alpha} \;,
\end{eqnarray*}
where the $f^{(a)}_{ABCD}\sim bv$
are related to the detailed form of the backscattering terms, 
and repeated  indices are summed.  

In the hydrodynamic equations for the classical values of these
fields, we expect damping terms similar to that in
Eq.~(\ref{precession2}).  In fact, there are two distinct sorts of
damping processes which need be considered.  First, like for the
flavorless problem, backscattering terms lead to decay of the
(non-chiral) $SU(4)$  currents, $\vec{J}^{A\alpha B\beta} =
v(\vec{J}^{A\alpha B\beta}_R -\vec{J}^{A\alpha B\beta}_L)$, which can
be included via a lifetime $\tau_J$.  Second, however, unlike in the
$SU(2)$ case, the flavor densities themselves can decay via
backscattering.  Only the true spin magnetization,
\[
\vec{M}=
(\vec{J}^{A\alpha A\beta}_R +\vec{J}^{A\alpha
  A\beta}_L)\vec\sigma_{\alpha\beta}/2 \;,
\]
and charge density,
\[
n=-e(\vec{J}^{A\alpha A\alpha}_R +\vec{J}^{A\alpha A\alpha}_L) \;,
\]
are
required to obey continuity equations, namely 
Eq.~(\ref{sum1}) and the usual
charge continuity equation, by spin-rotational and $U(1)$ invariance.
The remaining orthogonal linear combinations of the spin-flavor
densities can themselves decay, with some ``flavor decay rate''
$1/\tau_f \sim 1/\tau_J$.  For sufficiently long nanotubes and low
energies, $V\tau_f \ll 1$, these flavor densities become
negligible between contacts, and we expect to be able to simply ignore 
the flavor currents.  Given the smallness of the backscattering
couplings in nanotubes, however, this may occur only at very low
voltages and temperatures, and for very long tubes.  

A proper treatment of these effects at intermediate length scales is
technically rather complicated, and beyond the scope of this paper.
Nevertheless, the extension is in principle straightforwardly based on 
the techniques developed here.  It is amusing to note that the issue
of flavor currents is actually a concern even when all the leads are
ordinary paramagnetic metals, since even such contacts generically do
not respect flavor.

\subsection{Spin-flip scattering}

Up to this point, we have assumed that the QW in consideration is
itself spin-rotationally invariant.  In general, quantum mechanical
spin-orbit coupling mixes spin and orbital angular momentum, leading
to a small violation of spin conservation.  This can occur both as a
bulk and a boundary effect.  For semiconductor QWs, the spin-orbit
effects are well understood.  For SWNTs, they are expected to be
extremely small.  Indeed, in an ideal flat sheet of graphene,
spin-orbit effects are negligible due to the high symmetry of the
zone-boundary wavevector, the $p^z$ nature of the electronically
active carbon orbitals, and the small atomic number of carbon.  From a
simple tight-binding treatment (not shown here),
we indeed find a vanishing effect for ideal
graphene. In SWNTs,  bulk spin-orbit effects can then in principle
solely occur due to the curvature of the nanotube, to phonon distortions, and
to defects in the tube.  The first two factors are probably negligible
as they are suppressed both by the smallness of the relativistic
nature of spin-orbit coupling and by the smallness of the
nanotube curvature and the electron-phonon  coupling,
respectively.  Tube defects generally destroy the local symmetry of
the lattice and allow some spin-orbit scattering.  However, these same
defects also elastically scatter ordinary momentum, so that one may
estimate the spin-orbit scattering rate as
 $1/\tau_{\rm so} = \epsilon/\tau_J^{\rm el}$,
where $\epsilon \ll 1$ reflects the relativistic nature of the
microscopic spin-orbit coupling.  Since the elastic mean-free-path of
SWNTs is known to be of the order of microns, the
corresponding spin-orbit scattering length must be orders of magnitude
larger, and hence also of no importance for current tubes which are at
best a few tens of microns long.  Parenthetically, we note that in
multi-wall nanotubes,
 disorder is more important, and hence spin-orbit
scattering may be of more relevance.
In any case, it is straightforward to include the effect of spin
non-conservation theoretically by modifying Eq.~(\ref{sum1}) to
\begin{equation} \label{sum1a}
  \partial_t \vec{M} + \partial_x \vec{J} =  - \vec{M}/\tau_{\rm so} -\mu_B
  \vec{M}\times \vec{B}\;.
\end{equation}
Probably more significant is spin-flip scattering at the boundaries of
the QW and contacts, which are often much more disordered than the
bulk of the QW.  Such processes can be incorporated by a 
renormalization of the contact parameters $P, G,$ and $\vartheta$,
see Ref.~\onlinecite{brataas}.

\subsection{Dipolar fields}

A uniform magnetic field has already been included in
Eq.~(\ref{Hmag}).  In general, magnetic contacts give rise to a
spatially-varying dipolar field acting on the  electron
spin in the QW.   Provided the variation of this field is smooth on
the scale of the Fermi wavelength of the QW, however, the hydrodynamic
treatment of the magnetic field can be applied to this case as well,
simply letting $B \rightarrow B(x)$ in Eqs.~(\ref{sum1})
and (\ref{precession2}).
Because the characteristic spatial scale of variation of the magnetic
field is the size of the leads themselves, this condition should be
amply satisfied for nanotubes and most semiconductor QWs.  

One can get an idea of the magnitude of the effect of dipolar fields
by considering an idealized uniformly polarized spherical
FM of radius $r_c$ and magnetization 
$\vec{m}_0$  end-contacting the QW at $x=0$.
The external magnetic field of such a sphere is a pure
magnetic dipole, hence $B \sim |\vec{m}_0|/(1+x/r_c)^3$, with the usual
dipolar dependence on the orientation of $\vec{m}_0$.
To maximize the effects of the dipolar interactions, assume an Fe
contact (which has larger magnetization than Co or Ni), and a radius
$r_c$ comparable to the length $L$ of the QW, so that $B
\sim |\vec{m}_0|$ over the whole length.  For low-temperature iron,
 $|\vec{m}_0|
\approx 0.17$~T, and thus Eq.~(\ref{precnew})
leads to the typical phase change, estimated for a SWNT, of
\begin{equation}
  {{\gamma_{\rm dipolar}^{\rm Fe}} \over {2\pi}} \approx .003 \left(
    {L \over {1\mu}}\right) \;.
\end{equation}
Thus the effect of dipolar fields is probably negligible.

\section{Discussion}
\label{sec:outlook}

We conclude this paper by establishing a connection
to Andreev currents in superconductor-normal-superconductor (SNS)
junctions and by  pointing out some open questions.

\subsection{Analogies to Superconductors}
\label{sec:sns}

An interesting view on many of the
above results follows through an
analogy to Andreev processes in ballistic
SNS junctions. Consider for example the
device depicted in Fig.~\ref{fig0}.  Without loss of
generality, we choose the magnetization of the left FM lead as 
$\hat{m}_1=\hat{x}$, and that of the right lead in the $x-y$
plane, $\hat{m}_2 = \cos(\theta) \hat{x} - \sin(\theta) \hat{y}$.
Neglecting electron tunneling, magnetic fields and backscattering,
 the Hamiltonian 
fully decouples into spin and charge
components.  Because the spin Hamiltonian is independent 
of electron-electron interactions,
we are free to model it using effectively  non-interacting
Fermions $\psi$.  Note that this in no way implies that the QW is
non-interacting, but simply represents the physics of spin-charge
separation.  With boundary exchange couplings $K_1$ and $K_2$, we thus have
\begin{eqnarray}
  H_\sigma & = & -iv \int_0^L \! dx \, \psi^\dagger \tau^z\partial_x
  \psi^{\vphantom\dagger} - {K_1 \over 2} \left( \psi^\dagger_\uparrow(0)
    \psi^{\vphantom\dagger}_\downarrow (0) + {\rm
      h.c.}\right) \nonumber \\ 
  & & - {K_2 \over 2} \left( e^{i\theta} \psi^\dagger_\uparrow(L)
    \psi^{\vphantom\dagger}_\downarrow (L) + {\rm
      h.c.}\right)\;. \label{FLF} 
\end{eqnarray}
Equation (\ref{FLF}) must be supplemented by 
the boundary conditions $\psi_R 
= \psi_L$ at $x=0$ and $x=L$.  Consider now the
(spin-down) particle-hole transformation,
\begin{equation}
  \tilde\psi_\uparrow^{\vphantom\dagger} =
  \psi_\uparrow^{\vphantom\dagger}
\;, \qquad
  \tilde\psi_\downarrow^{\vphantom\dagger} =
  \psi_\downarrow^\dagger \;, 
\end{equation}
which retains canonical anticommutators for $\tilde\psi$ and preserves 
the boundary conditions.  Under this transformation, the kinetic terms 
in $H_\sigma$ are invariant, but the boundary terms become anomalous,
\begin{eqnarray}\label{SNS}
  H_\sigma & = & -iv \int_0^L \! dx \, \tilde\psi^\dagger
  \tau^z\partial_x 
  \tilde\psi^{\vphantom\dagger} \\ \nonumber 
  & & - \left(K_1 \tilde\Delta(0) + K_2
    e^{i\theta}\tilde\Delta(L) + {\rm h.c.}\right) \;,
\end{eqnarray}
where the pair field is 
\begin{equation}
  \tilde\Delta = {1 \over 2}\left( \tilde\psi^\dagger_{R\uparrow}
  \tilde\psi^\dagger_{L\downarrow} - \tilde\psi^\dagger_{R\downarrow}
  \tilde\psi^\dagger_{L\uparrow} \right) \;, 
\end{equation}
and we used the boundary conditions to remove the factor of $2$ in the 
magnetic exchange.  Equation (\ref{SNS}) is the Bogoliubov-deGennes
Hamiltonian for an SNS junction in the limit of large normal
reflection.  The latter limit is implied by the boundary conditions
$\tilde\psi_R=\tilde\psi_L$ at the ends.

  The presence of the pair-field
terms leads to Andreev reflection at the boundaries.  As is
well-known, such an SNS junction carries an {\sl equilibrium current}\
$\tilde{I}$
for any $\theta$ which is not a multiple of $2\pi$.  In particular, we 
expect\cite{SNSref} for $\theta$ of order one, 
\begin{equation} \label{apred}
  \tilde{I} = e v \left\langle \tilde\psi^\dagger
    \tau^z\tilde\psi^{\vphantom\dagger}\right\rangle \sim \left({K_1
      K_2 \over v^2}\right) {e v \over L} \;,
\end{equation}
in the fully coherent limit, $eV , k_B T \ll 
v/L$.  To translate this result back into the
spin problem, we can make a dictionary relating quantities in 
the two pictures.  Some interesting variables in the spin problem are
\begin{eqnarray*}
  J^z  =  {v \over 2} \psi^\dagger \tau^z\sigma^z
  \psi^{\vphantom\dagger} & = & \tilde{I}/2e\;, \\
  m^z = {1 \over 2} \psi^\dagger \sigma^z\psi^{\vphantom\dagger} & = & 
  {1 \over 2} : \tilde\psi^\dagger \tilde\psi: = \tilde{n}/2 \;, \\
  J^+ = v \psi^\dagger_\uparrow \tau^z
  \psi_\downarrow^{\vphantom\dagger} & = &
  v(\tilde\psi^\dagger_{R\uparrow} \tilde\psi_{R\downarrow}^\dagger -
  \tilde\psi^\dagger_{L\uparrow} \tilde\psi_{L\downarrow}^\dagger) \;, \\
  m^+ = {1 \over 2} \psi^\dagger_\uparrow
  \psi^{\vphantom\dagger}_\downarrow & = &
  \tilde\psi^\dagger_{R\uparrow} \tilde\psi_{R\downarrow}^\dagger + 
  \tilde\psi^\dagger_{L\uparrow} \tilde\psi_{L\downarrow}^\dagger \;.
\end{eqnarray*}
We see that $J^z$ and $m^z$ correspond to the charge current and density, 
respectively, in the transformed variables.  Thus the FM-LL-FM device indeed
carries a non-vanishing $z$-axis spin current.  Note, however, that
the in-plane magnetization corresponds to strange ``large-momentum''
pair fields in the analog SNS system.

For comparison to the results of the previous sections, note that our
hydrodynamic treatment gives zero spin current at zero applied
voltage.  This is not inconsistent, because Eq.~(\ref{apred}),
which is 
exact in equilibrium, predicts a spin current $J^z \sim v/L$ that
vanishes in the thermodynamic limit.  More precisely, the hydrodynamic 
results require incoherent transport, which holds e.g.~for
  $eV \gg v/L$.  
The hydrodynamic approach predicts $J^z \sim GP V$,
which can be crudely matched to the ``Andreev''
prediction (\ref{apred}).  In particular,
the hydrodynamic and
``Andreev'' currents are comparable when $eV \sim v/ PL$, that is
essentially at the boundary between the coherent and incoherent
regimes.  One learns from the SNS mapping that the spin current is
actually enhanced by coherence.  

In an SNS junction, one expects a proximity-effect induced pair field
within the normal region.  Naively one might therefore expect some
uniform  bulk magnetization in the $x-y$ plane.  This conclusion
is, however, false, as can be seen by rewriting the superconducting
pair field,
\begin{equation}
  \tilde\Delta = \psi_{R\uparrow}^\dagger
  \psi_{L\downarrow}^{\vphantom\dagger} + \psi_{L\uparrow}^\dagger
  \psi_{R\downarrow}^{\vphantom\dagger} \;.  \label{2kfmag}
\end{equation}
The pair field thus maps back to the $2k_F$
oscillatory component of the $x-y$ magnetization,
$\tilde\Delta = m^+_{2k_F} + m^-_{2 k_F}$,
but not to the uniform one.

\subsection{Outlook and open questions}

Let us finally summarize some of the 
open questions from our point of view,
and provide an outlook.

One  rather obvious
concern might be the {\sl incoherent}\ nature of transport
assumed in our study.  For sufficiently long quantum wires
and/or low conductance of the contacts, it certainly is
appropriate to assume a two-step sequential transport
mechanism through the FM-LL-FM  device.  What happens
if one has {\sl coherent}\ transport?  The latter 
situation could arise for higher-transparency contacts
or at very low energies.  However, from the analogy to
SNS junctions, we expect that our main conclusions are
qualitatively unaffected by coherence, and we therefore
do not expect a dramatic change.   Nevertheless, it would
be interesting to study this question in detail.  For
non-interacting electrons, this could be done in
the framework of a Landauer-type approach, see also
Refs.~\onlinecite{brataas} and \onlinecite{slonc}.

A related issue concerns the role of {\sl charging effects}\cite{grabert}
where transport through the LL is hindered by Coulomb blockade.
For low-transparency contacts,  these effects are known to be crucial
at energy scales below the charging energy $E_c$, which can be estimated for
a SWNT of length $L$, radius $R$, and background dielectric constant $\kappa$, 
\[
E_c= (e^2/\kappa L) \ln(L/R) \;,
\] 
and typically is of the order of a few meV.  Charging effects are
washed out by intermediate-to-high temperatures, and could in
principle be avoided altogether by using higher-transparency contacts
and/or long tubes.  We note that charging effects also tend to destroy
spin accumulation,\cite{charg} and therefore one has to be careful
that they are not present when experimentally testing for spin-charge
separation.  However, since they manifest themselves through quite
pronounced dependencies on external gate voltages, this issue is not
expected to create serious difficulties in practice.  Furthermore,
although our present theory does not include charging effects, this
could be accounted for easily via a proper treatment of the zero modes
in the bosonized version of the Luttinger liquid.\cite{book,moriond}

A very interesting extension of the methods of this paper is to
problems involving {\sl mesoscopic}
ferromagnetic contacts which are sufficiently small
so that their magnetization becomes {\sl dynamical}.  
Here the quantum wire/nanotube would mediate an effective
RKKY-type interaction between the FM magnetizations,
and novel transport phenomena can be anticipated.

It would now clearly
be of great interest to experimentally study the 
scenario put forward here.  
The probably best candidates for such experiments are
single-wall carbon nanotubes, which should offer the
unique possibility of observing spin-charge separation directly
on a single 1D quantum wire. 
In addition, the effects of backscattering were
shown to imply rather dramatic consequences
for spin transport, such as a sawtooth-like oscillatory
current-voltage relation.  Such spectacular consequences
of the electron-electron interactions have not been
predicted previously, but should be observable for 
long nanotubes at very low temperatures.

Future work should also address in detail the 2D generalization of
these ideas, which seems particularly interesting in the context of
some theories of high-$T_c$ superconductivity.  We hope that our paper
has convinced the reader that spin transport in strongly correlated
mesoscopic systems represents an exciting area of research that leads
to both fundamental insights and technologically useful devices.

\acknowledgments
R.E.~acknowledges support by the DFG under the Heisenberg and the Gerhard-Hess
program.  L.B. was supported by NSF grant DMR--9985255, and the
Sloan foundation.

\appendix
\section{Transport in a magnetic field}
\label{appendixa}

Here we outline the main step in the derivation of
the $I-V$ characteristics in a magnetic field for $b=0$,
see Sec.~\ref{sec:bb}.  To do so, 
we first eliminate $\vec{J}_{1,2}$ and $\vec{h}_2$ from
spin current conservation using the spin currents in
Eqs.~(\ref{j1}) and  (\ref{j2}), 
and the relations (\ref{hh2}) and (\ref{jj2}).
We are then left with the following bulky
relation determining $\vec{h}_1$,
\begin{eqnarray*} 
&& \frac{\vartheta}{4\pi}
 \cos\gamma \, (\hat{m}_1+\hat{m}_2)\times \vec{h}_1 + \frac{\vartheta}
{4\pi}(1-\cos\gamma) \Bigl [ (\hat{B}\cdot\vec{h}_1) (\hat{m}_2\times \hat{B})
\\ \nonumber
&& - \{ \vec{h}_1\cdot(\hat{m}_1\times\hat{B})\} \hat{B} \Bigr]
+ (1+G_\alpha^2) \sin\gamma \, \vec{h}_1\times \hat{B} 
\\ && + (\vartheta/4\pi)^2 \sin\gamma \,
\Bigl[ (\hat{B}\cdot\vec{h}_1) (\hat{m}_1\times \hat{m}_2)
+ (\hat{B}\cdot\hat{m}_1) (\hat{m}_2\times \vec{h}_1) \Bigr] \\
&&
+ \frac{\vartheta}{4\pi} G_\alpha \sin \gamma \Bigl[
\vec{h}_1 \{\hat{B}\cdot(\hat{m}_1+\hat{m}_2) \} -
(\hat{B}\cdot\vec{h}_1) \hat{m}_1 \\  && -
(\vec{h}_1\cdot\hat{m}_2)\hat{B} \Bigr]
+ 2 G_\alpha \Bigl[\cos\gamma \,\vec{h}_1
+ (1-\cos\gamma) (\hat{B}\cdot \vec{h}_1) \hat{B} \Bigr] \\  &&
= P I_\alpha \Bigl\{
\cos\gamma \, \hat{m}_1 -\hat{m}_2 + (1-\cos\gamma) (\hat{B}\cdot\hat{m}_1)
\hat{B} \\  && + \frac{\vartheta}{4\pi}
  \sin\gamma \, [(\hat{B}\cdot\hat{m}_2)\hat{m}_1 - \cos\theta \,\hat{B}
] + G_\alpha \sin\gamma\, \hat{m}_1\times \hat{B} \Bigr\}
\;.
\end{eqnarray*}
This equation is then analyzed for $\theta=\pi$ 
and special choices for $\vec{B}$ in Sec.~\ref{sec:bb}.

\section{Dissipationless precession} 
\label{appendixb}

The algebraic manipulations necessary to obtain the self-consistency 
equation (\ref{scs1}) in Sec.~\ref{sec:cc} are provided in this appendix.
With $(\hat{h}_1-\hat{h}_2)\cdot \hat{J}=0$, we can write
\begin{equation}\label{h1h2}
\hat{h}_1\cdot \hat{h}_2 = \left[ 1-
 (\hat{h}\cdot \hat{J})^2 \right]
\cos(\Delta \varphi) + 
 (\hat{h}\cdot \hat{J})^2 \;.
\end{equation}
Since $\vec{h}_1\cdot \vec{J}_1=\vec{h}_2\cdot \vec{J}_2$ is
conserved, by multiplying Eq.~(\ref{j1}) by
$\vec{h}_1$ and Eq.~(\ref{j2}) by $\vec{h}_2$, and exploiting
Eq.~(\ref{currcons}), we get
$\hat{h}_{1,2} \cdot \hat{J} = 0$.
This in turn implies directly that we can relate $\hat{h}_2$ to 
$\hat{h}_1$.  With $\vec{h} = \vec{h}_1$, we obtain
\begin{equation} \label{h2}
\hat{h}_2= \cos(\Delta\varphi) \hat{h} - \sin(\Delta \varphi) \hat{h}\times 
\hat{J} \;.
\end{equation}
The unknown spin chemical potential $\vec{h}$ can then be obtained
from spin current conservation, $\vec{J}_1= \vec{J}_2$,
with the currents specified in Eqs.~(\ref{j1}) and (\ref{j2}).

Using the abbreviations (\ref{abb1}), $W=J/PI_\alpha >0$,
and $Z=\vartheta/(4\pi G_\alpha)$, the relation
$\vec{h}\cdot \vec{J}_1=0$ gives 
$\hat{h} \cdot \hat{m}_1 = X$.
Furthermore,  from Eqs.~(\ref{j1}) and (\ref{j2}),
\begin{equation} \label{mj}
\hat{m}_1\cdot \hat{J}_1=\hat{m}_2 \cdot \hat{J}_2=  (1-X^2)/W \;.
\end{equation}
Next we use that $\hat{h}\cdot \vec{J}_2=0$ implies 
\[ 
\hat{h}\cdot \hat{m}_2= 
XZ \sin(\Delta \varphi) \hat{m}\cdot \hat{J} - X\cos(\Delta \varphi) \;,
\]
where $\hat{m} \cdot \hat{J}$ is determined by Eq.~(\ref{mj}).
In addition, $\hat{m}_2\cdot \vec{J}_1= \hat{m}\cdot \vec{J}$, 
see Eq.~(\ref{mj}), gives
\begin{eqnarray*}
 \hat{h}\cdot (\hat{m}_1\times \hat{m}_2) &=& (2/XZ)
[\sin^2(\theta/2)-X^2\cos^2(\Delta \varphi/2)] \\
&+&  X \sin(\Delta \varphi) \hat{m}\cdot\hat{J} \;. 
\end{eqnarray*}
Finally, we may employ the relation
$\hat{m}_1\cdot \vec{J}_2=\hat{m}\cdot\vec{J}$,
which yields 
\begin{eqnarray} \nonumber
&&XZ \cos(\Delta \varphi) \hat{h}\cdot (\hat{m}_1\times
\hat{m}_2) \\ \nonumber && - XZ\sin(\Delta \varphi) (\hat{m}\cdot \hat{J})
\, \hat{h}\cdot(\hat{m}_1-\hat{m}_2) \\ \label{temp}
 && -2\sin^2(\theta/2)+2X^2 \cos^2(\Delta \varphi/2) \\
 \nonumber && - X \sin(\Delta \varphi) 
\hat{m}_1 \cdot (\hat{h}\times \hat{J}) = 0 \;.
\end{eqnarray}
Alternatively, one could use spin conservation
of $\vec{J}\cdot (\hat{m}_1\times \hat{m}_2)$, which produces
the same answer.
When simplifying Eq.~(\ref{temp}), it is helpful to use the
following relation,
\[
\hat{m}_1 \cdot (\hat{h}\times\hat{J})  =
-  ZX(1-X^2)/W = - ZX \hat{m}\cdot \hat{J} \;,
\]
which follows from $\vec{J}_1 \cdot \hat{J} = J$.
Straightforward algebra then leads to
the self-consistency equation Eq.~(\ref{scs1}).

\section{Spin Diffusion}
\label{appendixc}

In this appendix, the technical steps in the 
derivation of the $I-V$ characteristics in the
presence of spin diffusion, see Sec.~\ref{sec:dd}, are given.
The charge current can be written as Eq.~(\ref{curr0}),
but with a modified quantity $X$,
\begin{equation} \label{newxdef}
X^2 =  \frac{G_\alpha}{P I_\alpha} \vec{h}_1\cdot \hat{m}_1  \;.
\end{equation}
In addition, we use $C_T= PI_\alpha L/\sigma_s$, with
the spin conductivity (\ref{spincond}).
To compute the current, we first
need to express $\vec{h}_2$ in terms of $\vec{h}_1=\vec{h}$
via the steady state diffusion-precession relation (\ref{precdiff}).
For symmetry reasons, $\vec{h}_2{}^2 = \vec{h}_1{}^2$, since we consider
identical contacts.  That directly 
implies from Eqs.~(\ref{j1}) and (\ref{j2}) that 
\begin{equation} \label{hj1} 
\vec{h}\cdot \hat{J}  = - 
\vec{h}_2\cdot \hat{J} =  W C_T /2 \;,
\end{equation}
where we use again $W=J/P I_\alpha$.
With the precession phase $\Delta \varphi$ defined in Eq.~(\ref{phase})
we then get instead of Eq.~(\ref{h2}),
\begin{equation} \label{h2new}
\vec{h}_2= \cos(\Delta\varphi) \vec{h} - \sin(\Delta \varphi) \vec{h}\times 
\hat{J} - WC_T \cos^2(\Delta \varphi/2) \hat{J} \;.
\end{equation}
Combined with Eq.~(\ref{hj1}), this allows to express
$h^2$ in terms of $X^2$ alone,
\begin{equation}\label{hoh2}
(hG_\alpha/P I_\alpha)^2= X^2 - W^2 G_\alpha  C_T/2 P I_\alpha \;.
\end{equation}
Here $\vec{J}_1{}^2 = J^2$ yields with $Z=\vartheta/\pi G_\alpha$,
\begin{equation} \label{w2new}
W^2 = \frac{(1-X^2)(1+Z^2 X^2)}{1+ G_\alpha C_T(1+Z^2)/(2P I_\alpha)} \;.
\end{equation}
Then Eq.~(\ref{mj}) still holds. 

We now employ spin current conservation to 
obtain a closed nonlinear self-consistency equation for finding
$X^2$ and thereby the current-voltage relation.
With $Q^2 = (G_\alpha/P I_\alpha) \vec{h}\cdot \hat{m}_2$,
the relation
$\hat{m}_1\cdot \vec{J}_2=PI_\alpha (1-X^2)$
gives after some massaging
\begin{eqnarray*} 
&& X^2 - \frac{ZW G_\alpha C_T}{2 P I_\alpha}
\sin(\Delta \varphi) - 4 \sin^2(\theta/2)\sin^2(\Delta
\varphi/2) \\ 
&+& \Bigl [\cos(\Delta \varphi)+ \frac{Z}{W}(1-X^2)\sin(\Delta \varphi)
\\  &+& \frac{ G_\alpha C_T}{P I_\alpha}
\cos^2(\Delta \varphi/2)(1+Z^2)\Bigr ] Q^2 
\\  &-& \frac{G_\alpha C_T}{P I_\alpha} 
\cos^2(\Delta\varphi/2) \cos(\theta) (1+Z^2 X^2) =0 \;.
\end{eqnarray*}
The second relation allowing to eliminate $Q^2$ comes from
$\vec{h}\cdot \vec{J}_2 = WC_T J/2$, and reads
\begin{eqnarray*}
&& \left[ 1 + \frac{ZWG_\alpha C_T}{ 2 PI_\alpha}  
\cot(\Delta\varphi/2)\right ] Q^2  
+ \cos(\Delta \varphi) X^2 \\
&-& \frac{Z}{W} \sin(\Delta \varphi) (1-X^2) X^2 
\\ &-& \frac{G_\alpha C_T}{P I_\alpha} \left(
1+ \frac{G_\alpha C_T}{2P I_\alpha} \right) \cos^2(\Delta \varphi/2) W^2
\\ &+& 
\frac{ZW G_\alpha}{2P I_\alpha} \cot(\Delta \varphi/2) \Bigl[
X^2 + \{2 \sin^2(\Delta \varphi/2) \\
& -& \nonumber  \frac{G_\alpha C_T}{P I_\alpha}
\cos^2(\Delta\varphi/2)\} (1-X^2)
\Bigr] = 0\;.
\end{eqnarray*}
Eliminating $Q^2$ from these two relations gives the
self-consistency equation for $X^2$ for arbitary temperature  and
applied voltage. The solutions
to this equation give directly the current via Eq.~(\ref{curr0}). 
One checks easily that this reproduces the $T=0$ self-consistency 
equation (\ref{scs1}). 

We shall now evaluate the self-consistency equation in 
the spin-diffusion dominated regime characterized
by $T \gg T^*$, with the scale
$T^*$ defined in Eq.~(\ref{tstar}).  This temperature
results from $f(T)=Z G_\alpha C_T/ P I_\alpha = 1$ for 
$T=T^*$.  Since $f(T)\gg 1$ for $T\gg T^*$, the 
above equations can be drastically simplified in
this regime.

For $T\gg T^*$, the self-consistency equation is again
solved by the discrete values (\ref{special}) of the precession
phase $\Delta \varphi$ indexed by
the winding number $n$.  Since under these conditions,
from Eq.~(\ref{w2new}), the
precession phase can be written as
\[
\Delta \varphi=   \left(\frac{2P I_\alpha}{
G_\alpha C_T} \right)^{1/2}\, \frac{bLPI_\alpha}{v} 
 X \sqrt{1-X^2} \;,
\]
it is then straightforward to derive Eq.~(\ref{ivnew})
in Sec.~\ref{sec:dd}.

\end{multicols}

\begin{references}
\bibitem{foot1}
Away from $\theta=\pi$, assuming nonzero exchange and $\alpha>0$,
our results hold for arbitrary $P$.
Furthermore, for $k_B T\ll eV$,  only the
weaker condition $P\ll 1+\alpha$  has to be imposed.
\end{references}
\end{document}